\documentclass{aa}
\bibpunct{(}{)}{;}{a}{}{,}
\usepackage[varg]{txfonts}
\usepackage{natbib,twoopt}
\usepackage{graphicx}   
\usepackage{amsmath}    
\usepackage{amssymb}    
\usepackage[usenames,dvipsnames]{xcolor}
\usepackage{multirow}
\usepackage{amsfonts}
\usepackage{mathrsfs}
\usepackage{nccmath}

\usepackage{ulem}
\usepackage{float}
\usepackage{placeins}
\usepackage{stfloats}
\usepackage{booktabs}

\usepackage{newtxtext,newtxmath}

\usepackage{soul}

\begin{document}

\title{Atomic diffusion and turbulent mixing in solar-like stars:\\ Impact on the fundamental properties of FG-type stars}
\author{Nuno Moedas\inst{1,2}\thanks{nmoedas@astro.up.pt} \and Morgan Deal\inst{1} \and Diego Bossini\inst{1} \and Bernardo Campilho\inst{1,2}}

\institute{ Instituto de Astrofísica e Ciências do Espaço, Universidade do Porto, CAUP, Rua das Estrelas, PT4150-762 Porto, Portugal
\and Departamento de Física e Astronomia, Faculdade de Ciências da Universidade do Porto, Rua do Campo Alegre, s/n, PT4169-007 Porto, Portugal}

\date{Received XXX / Accepted YYY}

\abstract {Chemical composition is an important factor that affects stellar evolution. The element abundance on the stellar surface evolves along the lifetime of the star because of transport processes, including atomic diffusion. However, models of stars with masses higher than about 1.2~M$_\odot$ predict unrealistic variations at the stellar surface. This indicates the need for competing transport processes that are mostly  computationally expensive for large grids of stellar models.}
{The purpose of this study is to implement turbulent mixing in stellar models and assess the possibility of reproducing the effect of radiative accelerations with turbulent mixing for elements like iron in order to make the computation of large grids possible.}
{We computed stellar models with the Module for Experiments in Stellar Astrophysics code and assessed the effects of atomic diffusion (with radiative acceleration) in the presence of turbulent mixing. Starting from a turbulent mixing prescription already calibrated on helium surface abundances of F-type stars as a reference, we parametrised the effect of radiative accelerations on iron with a turbulent diffusion coefficient. Finally, we tested this parametrisation by modelling two F-type stars of the Kepler Legacy sample.}
{We found that, for iron, a parametrisation of turbulent mixing that simulates the effect of radiative acceleration is possible. This leads to an increase in the efficiency of the turbulent mixing to counteract the effect of gravitational settling. This approximation does not affect significantly the surface abundances of the other elements we studied, except for oxygen and calcium. We demonstrate that this  parametrisation has a negligible impact on the accuracy of the seismic properties inferred with these models. Moreover, turbulent mixing makes the computation of realistic F-type star models including the effect atomic diffusion possible. This leads to differences of about 10\% in the inferred ages compared to results obtained with models that neglect these processes.}
{The inclusion of turbulent mixing and atomic diffusion with radiative accelerations allows a more realistic characterisation of F-type stars. The parametrisation of the effect of radiative acceleration on iron opens the possibility to compute larger grids of stellar models in a reasonable amount of time, which is currently difficult when the different chemical transport mechanisms, especially radiative accelerations, are considered, although this parametrisation cannot simulate the evolution of abundances of all elements (e.g. calcium).}


\titlerunning{Atomic diffusion and turbulent mixing in solar-like stars}
\authorrunning{Nuno Moedas \& et al.}
\maketitle

\section{Introduction}\label{sec:Intro}

The precise and accurate determination of stellar fundamental properties (such as age, mass, and radius) through stellar modelling is needed in many astrophysics applications, from the characterisation of exoplanetary systems to the reconstruction of the evolution and chemical history of the Milky Way. For example, stellar models have been largely used to interpret the data obtained with the missions \textit{Kepler}/K2 \citep{kepler} and  the Transiting Exoplanet Survey Satellite (TESS) \citep{TESS} and to put constraints on the targeted stars. These instruments provided and still provide high-quality asteroseismic data with exquisite precision that will be further enriched by future space missions such as PLAnetary Transits and Oscillations
of stars (PLATO/ESA; \citealt{PLATO}). However, the current stellar models are still suffering from large uncertainties (especially on the age), which will be critical for the interpretation of these new high-quality data (e.g.  PLATO requires 10\% accuracy on the age of a star similar to the Sun). For this reason, it is crucial to improve the ways we model stars.

One of the main ingredients responsible for the uncertainties on stellar ages is the transport of chemical elements. This transport is driven by processes (microscopic and macroscopic) in competition that lead to a redistribution of chemical elements with important effects on the internal structure, the evolution, and the surface abundances of stars. One of these processes is atomic diffusion. This process is mainly driven by pressure, temperature, {and chemical} gradients {\citep{Thoul,Baturin2006}} and affects the distribution of chemical elements at the surface and inside stars.

Atomic diffusion mainly consists in the competition between two sub-processes. One is the gravitational settling that makes the chemical elements move towards the interior, {except for hydrogen which is moved to the surface} of the star. The other is the radiative acceleration that pushes some elements{, mainly the heavy ones,} towards the surface of stars due to a transfer of momentum between photons and ions. {Its efficiency depends on the mass and metallicity of the stars, as shown in  \cite{Deal2018}, among others, increasing with mass and decreasing with metallicity.} Although, gravitational settling alone had proven to be successful in predicting the surface abundances of low-mass stars \citep{Chaboyer2001,Salaris2001}, radiative accelerations need to be included in stellar models with a small surface convective zone (e.g. in solar-metallicity stars with an effective temperature higher than $\sim6000$~K, \citealt{michaud15}).
However, for stars more massive than the Sun, the predicted surface abundance variations are often larger than those{} observed {in clusters \citep[e.g.][]{gruyters14,Gruyters2016,Semenova2020}}, which indicates the need to include other competing transport mechanisms. It is then crucial to identify and model the transport processes in competition with atomic diffusion {\citep[e.g.][]{Eggenberger2010,Vick2010,Deal2020,dumont20}}.
Nevertheless, the identification and accurate modelling of all processes competing with atomic diffusion is still ongoing and will require a considerable amount of effort. These processes are either diffusive or advective. Assuming that the processes are fully diffusive, an alternative solution is to parametrise the efficiency of a turbulent mixing induced by the competing transport processes, using the surface abundances of stars in clusters as constraints \citep{Gruyters2013,Gruyters2016,Semenova2020}. Recently, \citet{Verma2019}{ (hereafter VSA19)} calibrated a prescription of turbulent mixing using the helium surface abundances {of three F-type stars of the \textit{Kepler} Legacy sample}. These abundances were derived from an asteroseismic analysis of the glitch induced by the helium second ionisation region.

The accurate inference of the fundamental stellar properties of stars observed by large surveys requires large grids of stellar models with an accurate transport of chemicals. However, including these processes (e.g. atomic diffusion with radiative accelerations) in large grids of stellar models is still computationally expensive. Solutions to this issue need to be found for future large surveys (e.g. PLATO). The first goal of this paper is to characterise the effects of atomic diffusion (with radiative accelerations) for solar-like oscillating stars and to quantify the variation in $[\rm Fe/H]$ from the main sequence (MS) to the red giant branch (RGB) bump. In addition, and because [Fe/H] (iron abundance) is the main chemical constraint used for the inference of stellar fundamental properties, we also address the difference with [M/H] (overall metallicity) throughout the evolution. Focusing on F-type stars, which are the most impacted by atomic diffusion, the second goal is to quantify the impact of turbulent mixing (taking as a reference the calibration of {VSA19}) on the surface abundances. Then we propose a parametrisation for the effect of radiative acceleration on the iron surface abundance with an enhanced turbulent diffusion coefficient in order to make the computation of models faster without losing the contribution of radiative acceleration. Finally, we validate these models with the inferences of the properties of two F-type \textit{Kepler} Legacy stars.

This article is structured as follows. In Sect. \ref{stel_models} we present the input physics of the models.  In Sect. \ref{results} we address the effects of atomic diffusion and discuss its impact on the surface abundance variations solar-like  MS stars. In Sect. \ref{sec:Turb_Mix} we quantify the effect of turbulent mixing and parametrise the effect of radiative acceleration of iron. Finally, we test the impact of these models on the stellar property inferences of two \textit{Kepler} Legacy stars in Sect. \ref{sec:stars_infer}, and we conclude in Sect. \ref{conclu}.

\section{Stellar models}\label{stel_models}

\subsection{Input physics}

The stellar models are computed with the Modules for Experiments in Stellar Astrophysics (MESA r12778) stellar evolution code \citep{Paxton2011,Paxton2013,Paxton2015,Paxton2018,Paxton2019}. The input physics summarised below are the same for all the models, except for the transport of chemical elements and the opacity tables. We adopt the solar heavy elements mixture given by \citet{Asplund2009}. We use OP\footnote{\url{http://cdsweb.u-strasbg.fr/topbase/TheOP.html}} monochromatic opacity tables \citep{Seaton2005} when radiative accelerations are taken into account and OPAL\footnote{\url{https://opalopacity.llnl.gov/}} opacity tables \citep{Iglesias1996} in the other cases. We use the OPAL2005 equation of state \citep{Rogers2002}.
For nuclear reactions, we use the NACRE reaction rates \citep{Angulo1999} except for $^{14}$N(p,$\gamma$)$^{15}$O \citep{Imbriani2005} and $^{12}$C($\alpha$,$\gamma$)$^{16}$O \citep{Kunz2002}. For the boundary condition at the stellar surface we use Krishna-Swamy atmosphere \citep{Krishna1966}. {For convection we follow the prescription of \cite{Cox1968}}.
In the presence of a convective core we implemented core overshoot following an exponential decay with a diffusion coefficient, as presented in \citet{Herwig2000},

\begin{equation}
    D_{\rm ov}=D_0\exp\left(-\frac{z}{fH_p}\right),
\end{equation}
where $D_0$ is the diffusion coefficient at the border of the convective unstable region,  $z$ is the distance from the boundary of the convective region, $H_p$ is the pressure scale height, and $f$ is the overshoot parameter set to $f=0.01$.
Different solar calibrations are performed depending on the input physics of the models. The different $\alpha_{\rm MLT}$ values and initial chemical compositions are given in Sect.~\ref{sec:stellar_models}.

\subsection{Atomic diffusion}
 
Atomic diffusion occurs during the whole evolution of stars, and mainly acts in radiative zones (since convective motions almost instantaneously homogenise the chemical composition). Its impact at the surface depends on the extension of the convective envelope since its efficiency decreases with depth. Hence, a more efficient transport is present when the surface convective zone is small. This particularly affects the MS stars, where the extension of the envelope mainly depends on the stellar mass (higher mass, smaller convective envelope). On the other hand, during the sub-giant  (SG) phase and the beginning of the RGB, the envelope starts to increase, reaches its greatest depth (known as the  first dredge-up), and almost restores the surface metallicity to its initial value.

The chemical evolution of an element $i$ in the stellar interior is described by the following equation:
\begin{multline}
\label{diff_eq}
     \rho \frac{\partial X_{i}}{\partial t} =  
     A_i m_p \left[\sum_{j}^{} (r_{ji} - r_{ij})\right] +  \frac{1}{r^2} \frac{\partial }{\partial r} \left[r^2 \rho D_\mathrm{{\mathrm{T}}} \frac{\partial X_{i}}{\partial r} \right] -\\  \frac{1}{r^2} \frac{\partial }{\partial r} \left[r^2 \rho v_i \right].
\end{multline}
The first term takes into consideration the nuclear reactions, where $r_{ij}$ is the reaction rate of the reaction that transforms element $i$ into $j$. The {second} term takes into consideration all macroscopic diffusive processes that act inside the star, which are in competition with atomic diffusion with $D_{\rm {{\mathrm{T}}}}$ the turbulent diffusion coefficient, {$r$  the radial coordinate, and $\rho$  the local density.}

The last term corresponds to the effects of atomic diffusion, where  $v_i$ is the diffusion velocity of element $i$ that, in the case of a trace element, can be expressed as
 \begin{multline}
 \label{v_diff}
    v_i=D_{i,p}\left[-\frac{\partial \ln X_i}{\partial r}+{k_T}\frac{ \partial {\ln}T}{\partial r} +\frac{{(}Z_i{+1)} m_p g}{2k_BT} + \frac{A_i m_p}{k_BT}(g_{\rm rad,i} - g)\right],
\end{multline}
where $A_i$ is the atomic mass of element $i$, $m_p$ is the proton mass, $T$ is the temperature, $D_{i,p}$ is the diffusion coefficient of element $i$ relative to protons, $k_T$ is the thermal diffusivity, $k_B$ is the Boltzmann constant, and $Z_i$ is the atomic charge of {the} element $i$. {The first and second terms represent the effect of} the chemical and temperature gradients, respectively.  The {third} term {represents the effect} of the electric field. The {last and dominant term represents the effect of the} pressure gradient {and} is decomposed into two main processes, the radiative accelerations ($g_{\rm rad,i}$) and gravitational settling ($g$).

Atomic diffusion in MESA is computed following the \citet{Thoul} method with diffusion coefficients computed from \citet{paquette86} (see \citealt{Paxton2011} for more details).
\cite{Paxton2015} improved the treatment of atomic diffusion by including the effects of radiative acceleration following the work of \citet{Hu2011}. Radiative accelerations are computed using a modified version of the OP package (OPCD, \citealt{Seaton2005}).
We followed the recommendations of \citet{campilho22} to set up all the options provided by MESA to control the modelling of atomic diffusion.

We computed models including atomic diffusion with and without radiative acceleration.
When including radiative acceleration, the Rosseland mean opacity is computed using the OP monochromatic opacity tables in \citet{Seaton2005} instead of the OPAL tables at a fixed heavy elements mixture. This ensures that the opacity profile, hence the internal structure, is consistent with the internal redistribution of heavy elements induced by radiative accelerations during the whole evolution.

\begin{table*}
        \caption[]{Summary of the different stellar parameters and input physics used in each grid. {In the `Atomic Diffusion" column, $g$ indicates that the models include atomic diffusion without radiative accelerations and $g+g_\mathrm{rad}$ indicates that the models include atomic diffusion with radiative accelerations. In the following column $\Delta M_0$ is the value of reference mass used to compute turbulent mixing; `None' means that we do not include turbulent mixing in the grid} }
        \resizebox{\textwidth}{!}{
        \begin{tabular}{cccccccccc}
        \hline \hline
    \multirow{2}{*}{Grid}&\multicolumn{2}{c}{Mass (M$_\odot$)}&
    \multicolumn{2}{c}{${\rm [Fe/H]_i}$ }&\multicolumn{2}{c}{$\Delta Y/\Delta Z$}&{Atomic} &  {$\Delta M_0$ (M$_\odot$)}& Opacity  \\
    \cmidrule(lr){2-3}\cmidrule(lr){4-5}\cmidrule(lr){6-7}
    & Range & Step& Range & Step& Range & Step& {Diffusion}  & &Table\\\hline
    A&[0.7;1.7]&0.02&-0.44; 0.06; 0.46&---&0.4;1.23;2.8&---&${g}$&None& OPAL\\
    B&[0.7;1.7]&0.1&0.06&---&1.23&---&${g}+{g_\mathrm{rad}}$&None&OP mono\\
    C1&[0.7;1.7]&0.1&-0.04; 0.06; 0.16&---&1.23&---&${g}+{g_\mathrm{rad}}$&$5\times10^{-4}$&OP mono\\
    {C2}&{1.2; 1.4} &{0.1}&{0.06}&{---}&{1.23}&{---}&${g}$&${5\times10^{-4}}$&{ OPAL}\\
    D1&[1.3;1.5]&0.025&[-0.1;0.2]&0.05&[0.2;4.0]&0.01 in Y &${g}$&Parametrised& OPAL\\
    D2&[1.3;1.5]&0.025&[-0.1;0.2]&0.05&[0.2;4.0]&0.01 in Y &No&None& OPAL\\
    {D3}&{[1.3;1.5]}&{0.025}&{[-0.1;0.2]}&{0.05}&{[0.2;4.0]}&{0.01 in Y} &{No}&{None}& {OPAL}\\
    {D4}&{[1.3;1.5]}&{0.025}&{[-0.1;0.2]}&{0.05}&{[0.2;4.0]}&{0.01 in Y} &${g}$&{None}& {OPAL}\\\hline

\end{tabular} }
        \label{table:stellar_grids}
\end{table*}

\subsection{Turbulent mixing}\label{sec:Dturb_models}

The origins of many processes in competition with atomic diffusion are still unknown. The main candidates are either diffusive (e.g. rotation-induced mixing; \citealt{palacios03,talon08,dumont20}) or advective (e.g. mass loss; \citealt{Vick2010}), or both. In this work we consider that the competing transport processes are diffusive and all their contributions can be approximated by a turbulent diffusion coefficient. This coefficient ($D_\mathrm{T}$) was implemented in  {MESA} following the prescription described in \citet{Richer2000}
\begin{equation}
    D_\mathrm{T}=\omega D(\rm{He})_0\left(\frac{\rho_0}{\rho}\right)^n,
    \label{eq:Dturb}
\end{equation}
where $\omega$ and $n$ are constants, $\rho_0$ and $D(\rm{He})_0$ are respectively the density and the diffusion coefficient of helium at a reference depth, and  D(\rm{He}) was computed following the analytical expression given by \cite{Richer2000}:
\begin{equation}
    D(\rm{He})=\frac{3.3\times10^{-15}T^{2.5}}{4\rho\ln{(1+1.{1}25\times10^{-16}T^3/\rho)}}.
    \label{eq:DHe}
\end{equation}

Previous studies used either a fixed envelope mass ($\Delta M_0$) or a fixed temperature ($T_0$) for the reference depth.
The temperature was used as reference point when turbulent mixing was calibrated on lithium surface abundances in Population II stars \citep[e.g.][]{richard02, richard05, deal21c} and on the surface abundances of stars in clusters \citep[e.g.][]{Gruyters2013,Gruyters2016,Semenova2020,dumont21}. In all of these cases $\omega$ and $n$ were set to $400$ and $3$, respectively. In these studies the reference temperature was calibrated between ${\log_{10}(T_0)=5.7}$ and 6.5, the value being dependent on the type of stars (i.e. lower temperatures of Population II stars than solar-like stars).
On the other hand, the mass was used as a reference point to calibrate the turbulent mixing on the surface abundances of F- and A-type stars, where $\omega$ and $n$ were set to $10^4$ and $4$, respectively \citep{Michaud2011a, Michaud2011b}. The reference point in mass was found to be $\Delta M_0\sim[1-2]10^{-6}~$M$_\odot$ for these stars.
{VSA19} performed a similar calibration on three \textit{Kepler} stars. The calibration was made in order to obtain surface helium abundances that fit the observed glitch in the oscillation spectra that was caused by the second ionisation zone of helium. They found a reference mass $\Delta M_0=5\times10^{-4}$ M$_\odot$, which is higher than for  A- and F-type stars. In this work we use  $\Delta M_0$ as reference instead of $T_0$ because, as \cite{Richer2000}  noted,   the surface abundances of elements other than lithium mainly depend on the envelope mass mixed by turbulent mixing. Using $T_0$ could lead to different envelope masses throughout the evolution.
Since the focus of this paper is solar-like oscillating MS stars, we decided to use the value calibrated by {VSA19} as a reference.

\subsection{Grids of stellar models}\label{sec:stellar_models}

We computed different grids of stellar models to quantify the effect of atomic diffusion and turbulent mixing on solar-like MS stars:
\begin{itemize}
    \item Grid A includes atomic diffusion without radiative accelerations. For this grid we computed models with a range of masses equally spaced, three initial metallicities, and three helium enrichment ratios, including the solar calibration ($[\rm M/H]_i=0.06$~dex and the helium-to-heavy element enrichment ratio $\Delta Y/\Delta Z=1.23$). The solar-calibrated values are $\alpha_\mathrm{MLT}=1.7106532$, $X_0=0.71843711$, $Y_0=0.26673452$, and $Z_0=0.01482837$.
    \item Grid B has the same input physics {as} grid A except for the inclusion of radiative accelerations and the Rosseland mean opacity computed with OP monochromatic tables instead of standard OPAL tables. The grid is restricted to the solar chemical composition. We used the same solar-calibrated input parameter as Grid A. We tested that is has a negligible impact on the models.
    \item Grids C1 and C2 are similar to Grid B with the additional inclusion of the effect of turbulent mixing following the calibration of {VSA19}. Grid C1 is computed with solar metallicity $\boldsymbol{\pm}0.1$~dex to allow a comparison with an optimisation method (see Section~\ref{AIMS_models}). Grid C2 does not have radiative acceleration and only includes  a few   models for comparison with the models of Grids B and C1. Because the turbulent mixing parametrisation has no impact on solar models (i.e. the reference depth of Eq.~\ref{eq:Dturb} is inside the surface convective zone of the Sun), we used the same  solar-calibrated input parameters as Grid A.
    \item Grids D1, D2, D3, and D4 are computed around the parameter space of KIC~2837475 and KIC~11253226. All grids are used to infer the fundamental properties of both stars. Grid D1 includes the turbulent mixing parametrised in Sect.~\ref{calib-dturb} and gravitational settling. Grid D2 and D3 include no transport except convection. The difference between these two grids is in the solar-calibrated values. Grid D2 uses the same as Grid A, while grid D3 includes the solar-calibrated value consistent with its input physics ($\alpha_\mathrm{MLT}=1.5908152$, $X_0=0.72914669$, $Y_0=0.25765492$, and $Z_0=0.01319839$). Finally, D4 includes atomic diffusion without radiative accelerations, without turbulent mixing, and with solar-calibrated values similar to Grid~A.
\end{itemize}
Grid A is used as a reference when compared to grids B and C1. For  grids C1, C2, D1, D2, D3, and D4 individual frequencies are computed using the GYRE oscillation code \citep{Townsend2013}. The parameters of the grids are summarised in Table~\ref{table:stellar_grids}.

\begin{figure}
    \centering
    \includegraphics[width=.95\columnwidth]{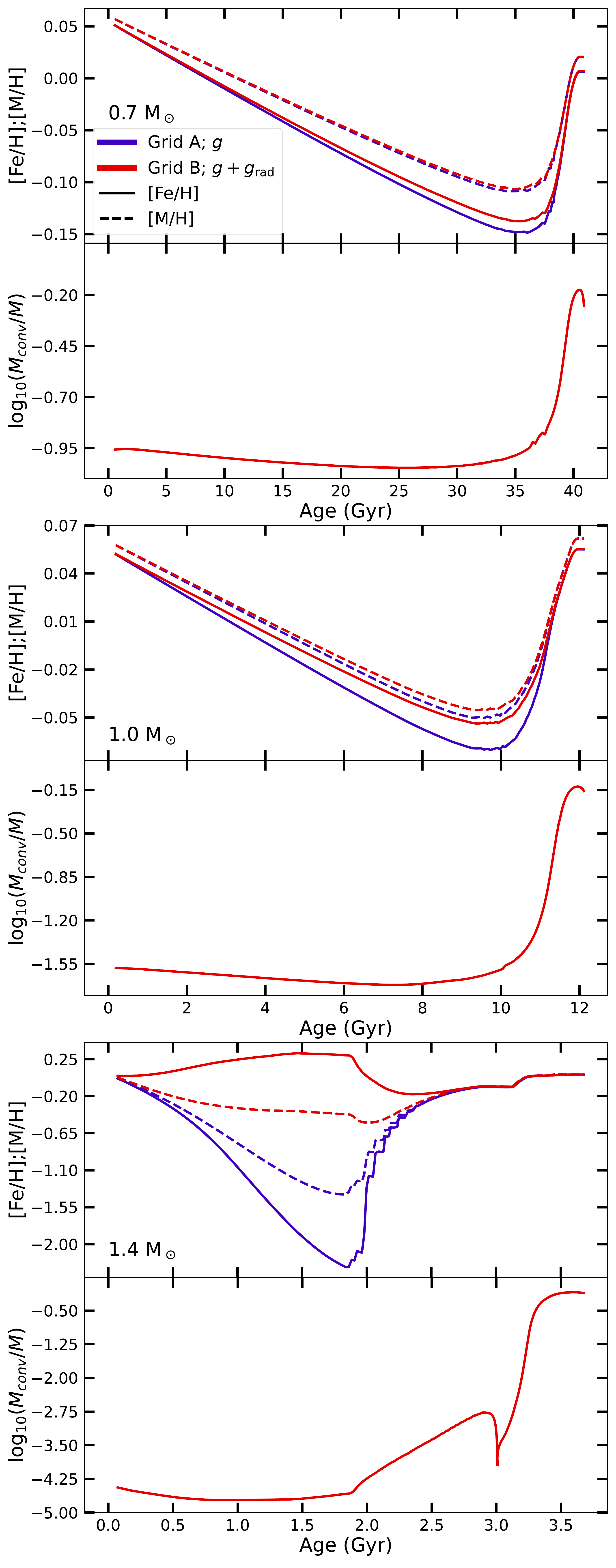}
    \caption{Variation in $\rm [Fe/H]$ (solid lines, top of all subplots) and [M/H] (dashed lines, top of all subplots), and evolution of the convective envelope mass divided by the total mass of the model (bottom of all subplots) from the ZAMS to the tip of the RGB for [Fe/H ]$\rm _i=0.06$. The red and blue lines represent models with and without radiative acceleration. The top panels are for 0.7~M$_\odot$ models, the middle panels for 1.0~M$_\odot$ models, and the bottom panels for 1.4~M$_\odot$ models.}
    \label{fig:ad_rad}
\end{figure}

\section{Atomic diffusion and surface abundances}
\label{results}

In this section, we focus on the evolution of the surface chemical composition, quantifying the differences between the predicted [Fe/H] and [M/H] when atomic diffusion is included in stellar models. We also compare the surface abundance evolution induced by atomic diffusion including or not radiative accelerations.

\subsection{Surface abundances}\label{ab}

{The chemical composition of a star can be defined by}
\begin{equation}
    {X+Y+Z=1,}
\end{equation}
where $X$ is the mass fraction of hydrogen, $Y$ is the mass fraction of helium ($^3$He + $^4$He), and $Z$ is the mass fraction of all the elements heavier than helium. However, these quantities are not directly observed in stars, but the surface abundances of individual element relative to other elements are. By definition, the element surface abundance of an element $A$ relative to an element $B$ is
\begin{equation}
    [A/B]=\log_{10}(N_A/N_B)-\log_{10}(N_A/N_B)_\odot,
\end{equation}
where $N_A$ and $N_B$ are the surface number fraction of the elements $A$ and $B$, respectively. The term indexed with $_\odot$ refers to the photospheric solar values. Following this definition, the iron abundance can be obtained with the expression
\begin{equation}
    {\rm[Fe/H]}=\log_{10}(N_{\rm Fe}/N_{\rm H})-\log_{10}(N_{\rm Fe}/N_{\rm H})_\odot.
    \label{eq:FeHfe2}
\end{equation}
The expression can be converted into mass fraction with the  expression
\begin{equation}
    \log_{10}(X_{\rm Fe}/X_{\rm H})_\odot =\log_{10}(N_{\rm Fe}/N_{\rm H})_\odot+\log_{10}(A_{\rm Fe})
    \label{eq:sunamb}
,\end{equation}
where $X_{\rm H}$ and $X_{\rm Fe}$ are the surface hydrogen and iron mass fractions, $N_{\rm H}$ and $N_{\rm Fe}$ are the number of atoms of hydrogen and iron at the solar surface, and $A_{\mathrm{Fe}}$ is the atomic mass of iron.
Most of the time, instead of using eq. \ref{eq:FeHfe2}, the surface iron abundance [Fe/H] is approximated by the metallicity [M/H] with the expression
\begin{equation}
    \rm[Fe/H]\sim{\rm[M/H]}=\log_{10}(Z/X)-\log_{10}(Z/X)_\odot.
    \label{eq:FeHz2}
\end{equation} 
To understand the possible uncertainties caused by this approximation for stars others than the current Sun, we compare the surface iron abundances obtained from  Eq.~\ref{eq:FeHfe2} with the metallicity obtained from Eq.~\ref{eq:FeHz2}.
For both equations, we use the solar values of \citet{Asplund2009}, $\log_{10}(N_{\rm Fe}/N_{\rm H})_\odot=-4.50$ and $\log_{10}(Z/X)_\odot=-1.7423$.

\subsection{Variation in [Fe/H] with  evolution}\label{sec:Fe_evo}

Figure~\ref{fig:ad_rad} shows the evolution of $\rm [Fe/H]$ estimated using  Eq.~\ref{eq:FeHfe2} (solid lines) and the metallicity estimated with Eq.~\ref{eq:FeHz2} (dashed lines) for three different masses, with and without radiative acceleration. In this section we first focus on the global behaviour of the iron surface abundance throughout the evolution.  \\

\begin{figure}
    \includegraphics[width=\columnwidth]{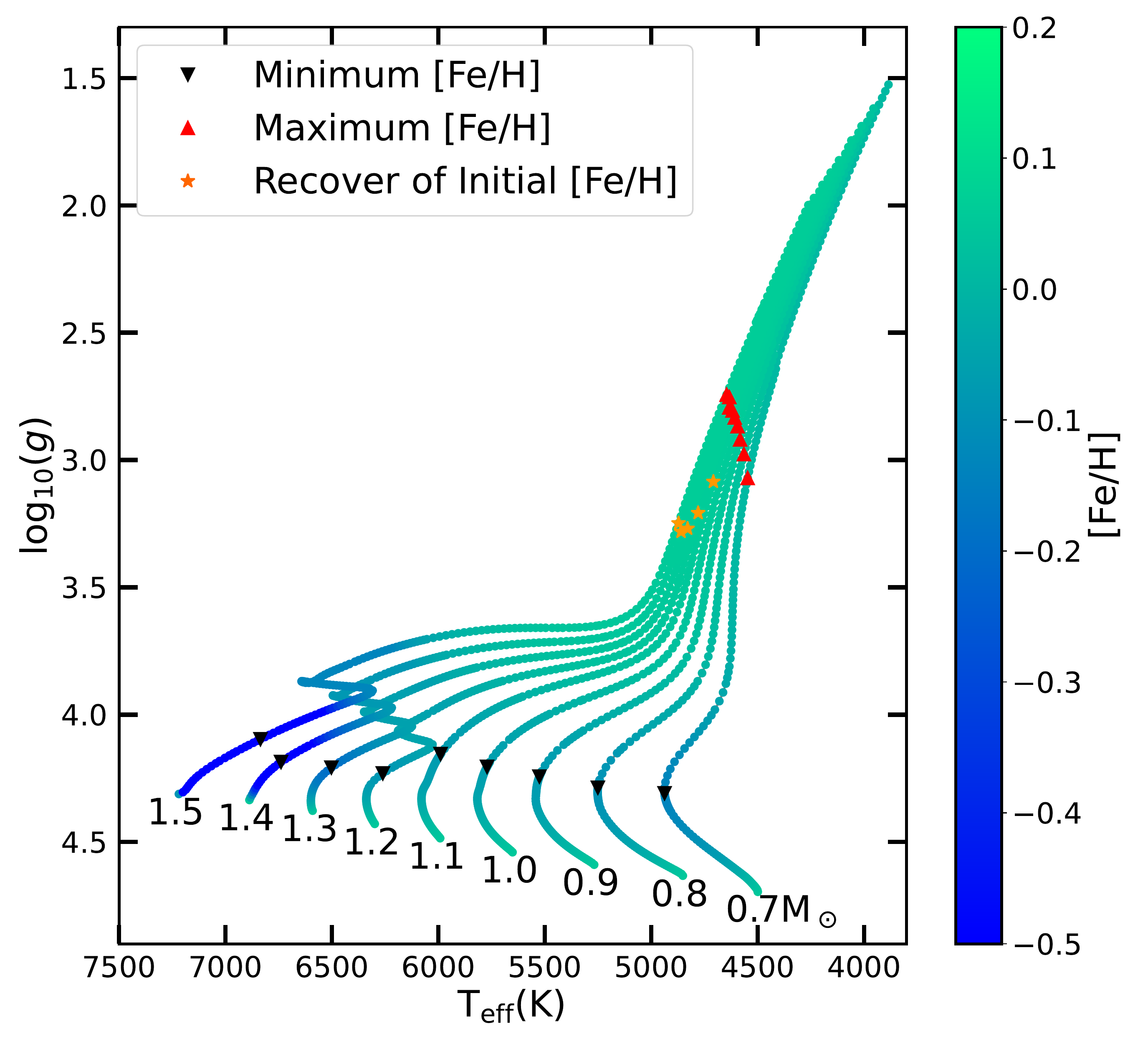}
    \caption{ Kiel diagram of models from Grid A with [Fe/H ]$\rm _i=0.06$ and $\frac{\Delta Y}{\Delta Z}=1.23$. The colour indicates the value of $\rm [Fe/H]$ at the stellar surface, the triangles down are the points where $\rm [Fe/H]$ reach a minimum, the triangles up are the points where $\rm [Fe/H]$ reaches a maximum, and the star symbols are the points where $\rm [Fe/H]$ reaches the initial value.}
    \label{fig:HR}
\end{figure}

\noindent\textbf{Models including atomic diffusion without radiative accelerations (grid A)}:

During the evolution of models without radiative acceleration, the surface $[\rm Fe/H]$ (blue solid curves) decreases until it reaches a minimum (largest depletion,   LD).
This minimum depends on the mass.
On one hand, the LD is larger for a 0.7~M$_\odot$ model than for a 1.0~M$_\odot$ model because the duration of the MS is longer for lower masses, so atomic diffusion has more time to act.
On the other hand, the LD is larger for a 1.4~M$_\odot$ model than for a 1.0~M$_\odot$ model because atomic diffusion is much more efficient for higher masses. This is partly due to the smaller size of the surface convective zone.
Figure \ref{fig:HR} shows the Kiel diagram (log$(g)$ against $T_{\rm eff}$) for some models of grid A (colour-coded for the value of $\rm [Fe/H])$. The LD of each track is represented by the down triangles. We can see in this figure that for models without convective cores, this point occurs at the end of the MS, while for models with convective cores, it occurs in the MS.

After the LD the surface $\rm [Fe/H]$ increases  to its maximum due to the first dredge-up (i.e. the penetration of the surface convective layers during the sub-giant branch and low RGB). This point in the diagram is represented by the up triangle in Fig.~\ref{fig:HR}.
The level of [Fe/H] reached depends on the stellar mass, with values close to the initial one. For the models presented in Fig.~\ref{fig:HR}, only those with masses higher than 1.0~M$\odot$ can reach or slightly surpass the initial composition.
 Figure~\ref{fig:ad_rad} shows, for the 1.4 M$_\odot$ models, that the iron abundance reaches a local maximum at the end of the MS, then decreases until the end of the MS hook, before increasing again. This phenomenon is due to the presence of a convective core in the MS. During the transition from MS to the SG phase, the stellar structure adjusts rapidly to the cessation of  nuclear reactions in the core, which induces a brief pause in the deepening of the surface convective envelope (around 3~Gyr in the bottom panel of Fig.~\ref{fig:ad_rad}).

After the maximum value of $\rm [Fe/H]$ is reached on the RGB phase, atomic diffusion slightly decreases it (up to about $10^{-5}$~dex).
However, the depletion is much slower due to the large extent of the convective envelope. 
These effects are insignificant compared to the observed uncertainties on [Fe/H]. A zoom-in of Fig.~\ref{fig:HR} around the  F-type stars is also presented in the  top panels of Fig.~\ref{fig:mdiagr}.
\\

\noindent\textbf{Models including atomic diffusion with radiative accelerations (grid B)}:

With radiative accelerations (red curves), the behaviour is similar for the 0.7 and 1.0~M$_\odot$ models. The iron surface abundance decreases with time, albeit at a lower rate, although for models with 1.4~M$_\odot$ we can see a substantial difference in the evolution of [Fe/H].
We see that the [Fe/H] increases at the surface during the MS evolution leading to higher abundances than the initial values, which avoids the large depletion of iron seen for models without radiative acceleration.
After the MS the values of [Fe/H] becomes similar to the models that do not include radiative accelerations and follow the same evolution. This shows that by including radiative accelerations the chemical evolution after the MS induces a negligible difference in the surface abundances for stars with masses lower than 1.4~M$_\odot$ at solar metallicity. All the results presented in this section are consistent with previous studies \cite[e.g.][and references therein]{Deal2018}.
The middle panels of Fig. \ref{fig:mdiagr} show this evolution of the surface [Fe/H] and [M/H] for F-type stars.

\subsection{[Fe/H] vs [M/H]}\label{sec:FeH_vs_MH}

Independently of the transport included in stellar models, it is important to compare the correct chemical indicators between data and models. For example, even if [M/H] can be approximated by [Fe/H] in a specific case (i.e. the Sun), this approximation is only valid if the ratio of $X(\rm Fe)$ to $Z$ remains the same. However, it has already been shown that this is not always the case, for example for alpha-enriched stars \citep{Salaris2001} or for F-type stars in which atomic diffusion modifies the chemical mixture \citep{Deal2018}. This is especially crucial when only [Fe/H] is used as a chemical constraint to infer the stellar properties of stars.

For the 0.7 and 1.0~M$_\odot$ models with and without radiative accelerations (grids B and A, respectively),  Fig.~\ref{fig:ad_rad} shows that both the surface $\rm [Fe/H]$, estimated with the iron abundance (eq. \ref{eq:FeHfe2}), and the actual metallicity, estimated using $Z/X$ ([M/H], eq. \ref{eq:FeHz2}) have a similar evolution, with the first having a faster decrease rate than the second. For the 1.4~M$_\odot$ model with radiative accelerations, we see instead that $\rm [Fe/H]$ increases while [M/H] decreases. This is in agreement with the results of \citet{Deal2018}. This occurs because the transport of chemical elements induced by atomic diffusion is different for each element. While iron is accumulated at the surface, most of the metals are depleted from the surface. Without radiative accelerations we expect iron to be depleted more quickly than the other metals.

Therefore, with radiative accelerations, approximating [Fe/H] using $Z/X$ induces an overestimation of the actual $\rm [Fe/H]$ for the low masses and an underestimation for the case of 1.4M$_\odot$ and higher masses. For models that do not include radiative accelerations the difference can be up to $\sim 0.04$~dex for a 0.7~M$_\odot$ model, $\sim$0.02~dex for a 1.0~M$_\odot$ model, and $\sim$0.8~dex for a 1.4~M$_\odot$ model, at solar metallicity. For models including radiative accelerations, this difference can be up to $\sim 0.03$, $\sim 0.01$, and $\sim 0.8${~dex} for the  0.7, 1.0, and 1.4~M$_\odot$ models, respectively. The effect is larger for the 1.4~M$_\odot$ model since the diffusion timescale is smaller {compared} to the lower masses, and the differences between $Z$ and iron are then more significant. This shows how crucial the definition of metallicity in stellar models may be, and how the way it is compared to abundances obtained from observations can lead to uncertainties, especially for the more massive stars in which competing transport to atomic diffusion is not always efficient (e.g. {AmFm} stars; \citealt{Richer2000}).

\subsection{Variation in helium surface abundances}

\begin{figure}
    \centering
    \includegraphics[width=\columnwidth]{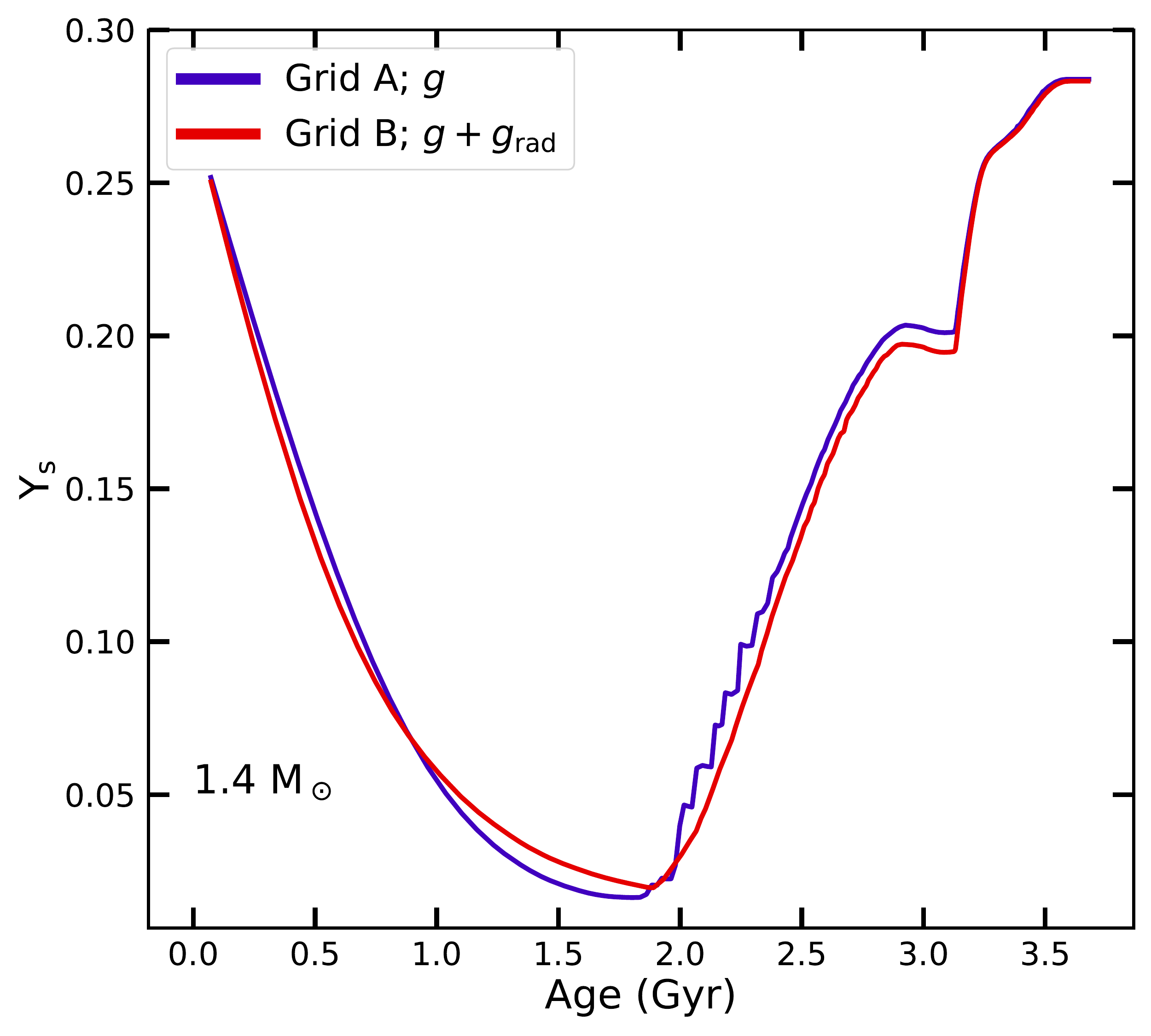}
    \caption{Evolution of the surface mass fraction of helium with (red curves) and without (blue curves) radiative accelerations for a 1.4M$_\odot$ stellar model with $\rm [M/H]_i=0.06$ and $\Delta Y/\Delta Z=1.23$.}
    \label{fig:rad_he}
\end{figure}

Atomic diffusion acts with different efficiency for each element. Not all the elements are in fact supported by radiative acceleration. This is the case of helium. As shown in Fig. \ref{fig:rad_he} for a 1.4~M$_\odot$ model, the helium surface abundance goes down to 0.01 in mass fraction, while it is not expected to go below about 0.18 for solar-like
oscillating MS stars \citep{Verma2019a}. This strengthens the need for the implementation of competing transport processes in stellar models.

\section{Turbulent mixing in stellar models}
\label{sec:Turb_Mix}

As seen in the previous section, atomic diffusion changes the surface chemical composition {and radiative accelerations need to be included for the more massive models}. However, the variations shown in Fig. \ref{fig:ad_rad} for the  1.4 M$_\odot$ models are unrealistic compared to the variations observed for stars in clusters \citep[e.g.][]{gruyters14,Gruyters2016,Semenova2020}. To account for the missing transport processes in stellar models, turbulent mixing was proposed as a solution \citep[e.g.][]{Richer2000,Michaud2011b}.
 In this section we use the expression proposed by \cite{Richer2000} for turbulent mixing, taking as a reference the calibration performed by {VSA19} and we quantify the impact of this calibration on the surface abundances of solar-like oscillating MS stars. Then we carry out a parametrisation of the turbulent mixing that takes into account the effects of radiative accelerations with the objective to make the computation of stellar models faster.
From here we focus on stars with convective cores for which atomic diffusion leads to unexpected surface abundance variations. This corresponds to stars with mass higher than $\sim$1.2~M$_\odot$ at solar metallicity. This mass is different for every initial chemical composition.

\subsection{Evolution of the surface abundances}

\begin{figure}
    \centering
    \includegraphics[width=\columnwidth]{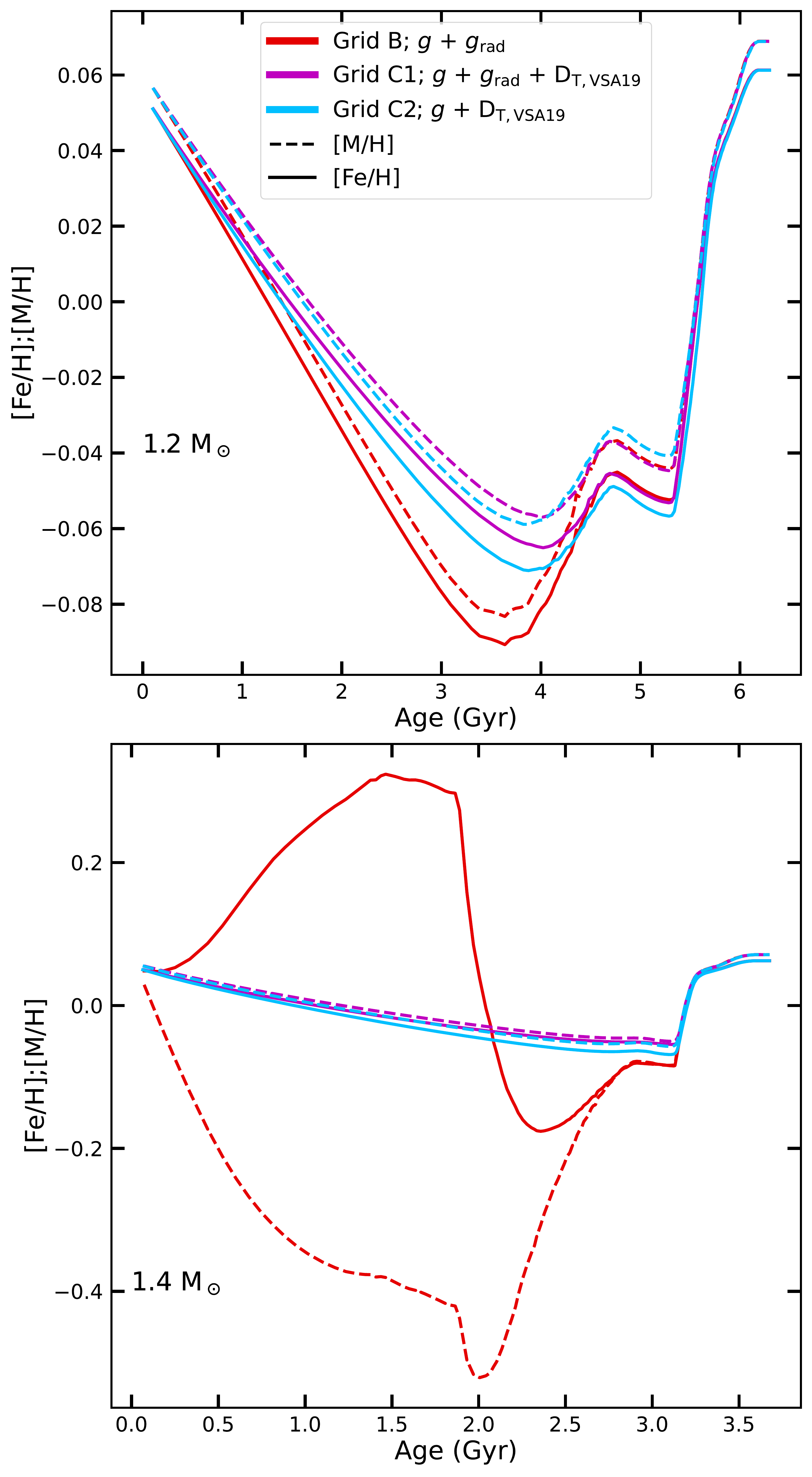}
    \caption{{Variation in [Fe/H] (solid lines) and [M/H] (dashed lines) from the ZAMS to the tip of the RGB for [Fe/H ]$\rm _i=0.06$. The purple and red lines represent models with and without turbulent mixing (both with radiative acceleration), and the cyan lines represent the model with only turbulent mixing. The top panel is for the 1.2~M$_\odot$ models and the bottom panel is for the 1.4~M$_\odot$ models.}}
    \label{fig:Dturb_Fe}
\end{figure}

We first investigate the surface abundance evolution of iron with models that include the effect of turbulent mixing, following the calibration in VSA19, and atomic diffusion with radiative accelerations. The evolution of the surface [Fe/H] for two different masses (1.2, 1.4~M$_\odot$) is shown in Fig. \ref{fig:Dturb_Fe}. We start by focusing on the models including radiative accelerations and turbulent mixing (grid C1, purple curves). The inclusion of turbulent mixing attenuates the depletion of [Fe/H] for the stellar model with 1.2 M$_\odot$. In this case the LD is about 0.02~dex smaller than the model without turbulent mixing (grid B, red curves). The inclusion of turbulent mixing has a more significant effect in the stellar model with 1.4 M$_\odot$.
The surface enrichment predicted by the model that includes atomic diffusion with radiative accelerations is now prevented. Instead, the turbulent mixing model exhibits a steady depletion of iron during evolution because the envelope mass homogenised by turbulent mixing reaches depths where the effects of gravitational settling are dominant compared to radiative acceleration.

For the models including turbulent mixing and atomic diffusion without radiative accelerations (grid C2, cyan curves) the depletion is larger than in the previous case (grid C1).
Even if radiative accelerations are not dominant compared to the gravity below the reference point ($\Delta M_0=5\times10^{-4}$~M$_\odot$, above this point the turbulent mixing homogenises the chemical composition), a difference between grid C1 and C2 models is still present (up to 0.015~dex for the 1.4~M$_\odot$ models). This indicates that radiative accelerations should still not be neglected in this case.

The difference between [M/H] and [Fe/H] is about 0.01~dex for the 1.2 and 1.4~M$_\odot$ models including turbulent mixing of grid C2. This is smaller than for the models presented in Sect.~\ref{sec:FeH_vs_MH}, but still close to the order of magnitude of the uncertainties of surface abundances sometimes found in the literature.

\begin{figure}
    \centering
    \includegraphics[width=\columnwidth]{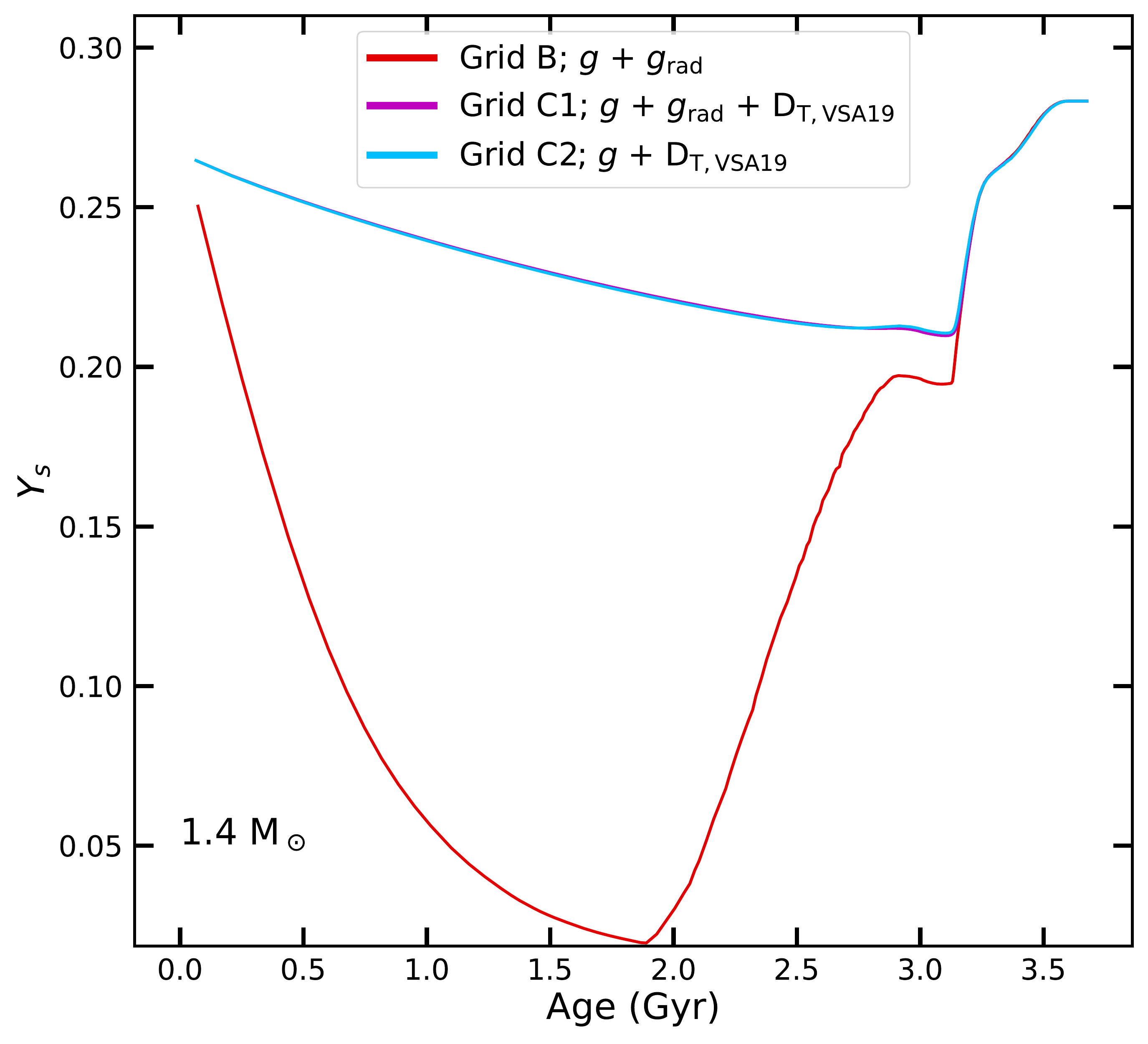}
    \caption{Same   as Fig.~\ref{fig:Dturb_Fe}, but  for the evolution of helium surface abundances of the 1.4~M$_\odot$ model.}
    \label{fig:Dturb_Ys}
\end{figure}

Lighter elements like helium are not affected by radiative accelerations. Models including  these processes or not will predict the same depletion of helium at the surface. This can be seen in Fig.~\ref{fig:Dturb_Ys}, where the helium surface abundance is shown for the 1.4~M$_\odot$ models presented in Fig.~\ref{fig:Dturb_Fe}. As expected for this element, models with turbulent mixing predict the same surface abundance variation with and without radiative acceleration. We can also see that at the beginning of the evolution (at the zero age main sequence, ZAMS) there is a different abundance at the stellar surface. This is due to an extra depletion of helium occurring during the PMS because of the more efficient   atomic diffusion in models without turbulent mixing.

\subsection{Parametrising radiative acceleration on iron with a turbulent diffusion coefficient}\label{calib-dturb}

As seen in the previous section, neglecting radiative accelerations when the turbulent mixing calibrated by VSA19 is taken into account leads to differences of up to 0.015~dex in [Fe/H]. This difference could even be larger for more massive stars or stars undergoing less efficient competing macroscopic transport. With the objective of computing large grids of stellar models for the fundamental stellar property inferences of large-scale surveys, the effect of radiative accelerations should be included somehow for accurate stellar properties. Since iron is depleted in all models including the VSA19 turbulent mixing considered in this study, it is possible to parametrise the effect of the radiative acceleration for this element by an increase in the efficiency of turbulent mixing (i.e. an increase in  the mass of the reference point $\Delta M_0$). This ensures that the computed models will have an iron surface abundance evolution that is close to the accurate value. Moreover, the implementation of turbulent mixing significantly reduces the variation in the metal mixture in the whole model, thus allowing the use of the classical opacity table. Simultaneously, it reduces the computationally demanding calculation of radiative accelerations and allows  larger grids to be built in shorter times.
Nevertheless, this parametrisation of the effect of radiative accelerations on iron should only be used in specific applications such as the stellar properties inference, which only uses [Fe/H] as a constraint for the chemical composition. An alternative and optimal solution would be to use the single valued parameter (SVP) approximation (\citealt{alecian20}) in order to compute radiative accelerations more efficiently. However, this method has not been implemented yet in MESA.

The parametrisation we propose is equivalent to rewriting the diffusion equation as 
\begin{multline}
    \label{param_diff}
     \rho \frac{\partial X_{i}}{\partial t} = 
     A_i m_p \left[\sum_{j}^{} (r_{ji} - r_{ij})\right] +  \frac{1}{r^2} \frac{\partial }{\partial r} \left[r^2 \rho D_\mathrm{T,Fe} \frac{\partial X_{i}}{\partial r} \right] -\\ \frac{1}{r^2} \frac{\partial }{\partial r} \left[r^2 \rho v_i' \right] ,
\end{multline}
with

\begin{equation}
 v_i'=D_{i,p}\left[-\frac{\partial \ln {X_i}}{\partial r} +{k_T}\frac{\partial {\ln}T}{\partial r}+\frac{{(}Z_i{+1)} m_p g}{2k_BT}  -\frac{A_im_pg}{k_BT}\right].
\end{equation}
\noindent where $D_\mathrm{T,Fe}$ is parametrised using eq. \ref{eq:Dturb} to make the surface abundance variation in iron match that  of models including radiative accelerations.

We quantify $\Delta M_0$ as a function of the stellar mass {following the procedure described in Appendix \ref{sec:calibr_dturb}}. Figure~\ref{fig:Dturb_call} shows the increase in $\Delta M_0$ needed to obtain a  [Fe/H] evolution similar to  models including turbulent mixing (${D_{\rm T,VSA19}}$) and radiative accelerations {(grid C1)}. The increase in $\Delta M_0$ with mass is explained by the fact that for more massive stars radiative accelerations are more efficient at pushing iron to the surface. For the input physics (see Section~\ref{sec:stellar_models}), the initial chemical composition (solar calibrated), and the reference calibration value ${D_{\rm T,VSA19}}$
used to compute the approximation and the models we derive a simple linear expression that describes the variation in $\Delta M_0$ with the mass:

\begin{equation}\label{eq:param}
    \Delta M_0 \left(\frac{M^\ast}{M_\odot}\right)= 3.1\times 10^{-4} \times \left(\frac{M^\ast}{M_\odot}\right) + 2.7\times 10^{-4}.
\end{equation}
\noindent This expression should be calibrated every time the input physics of the model changes.

\begin{figure}
    \centering
    \includegraphics[width=\columnwidth]{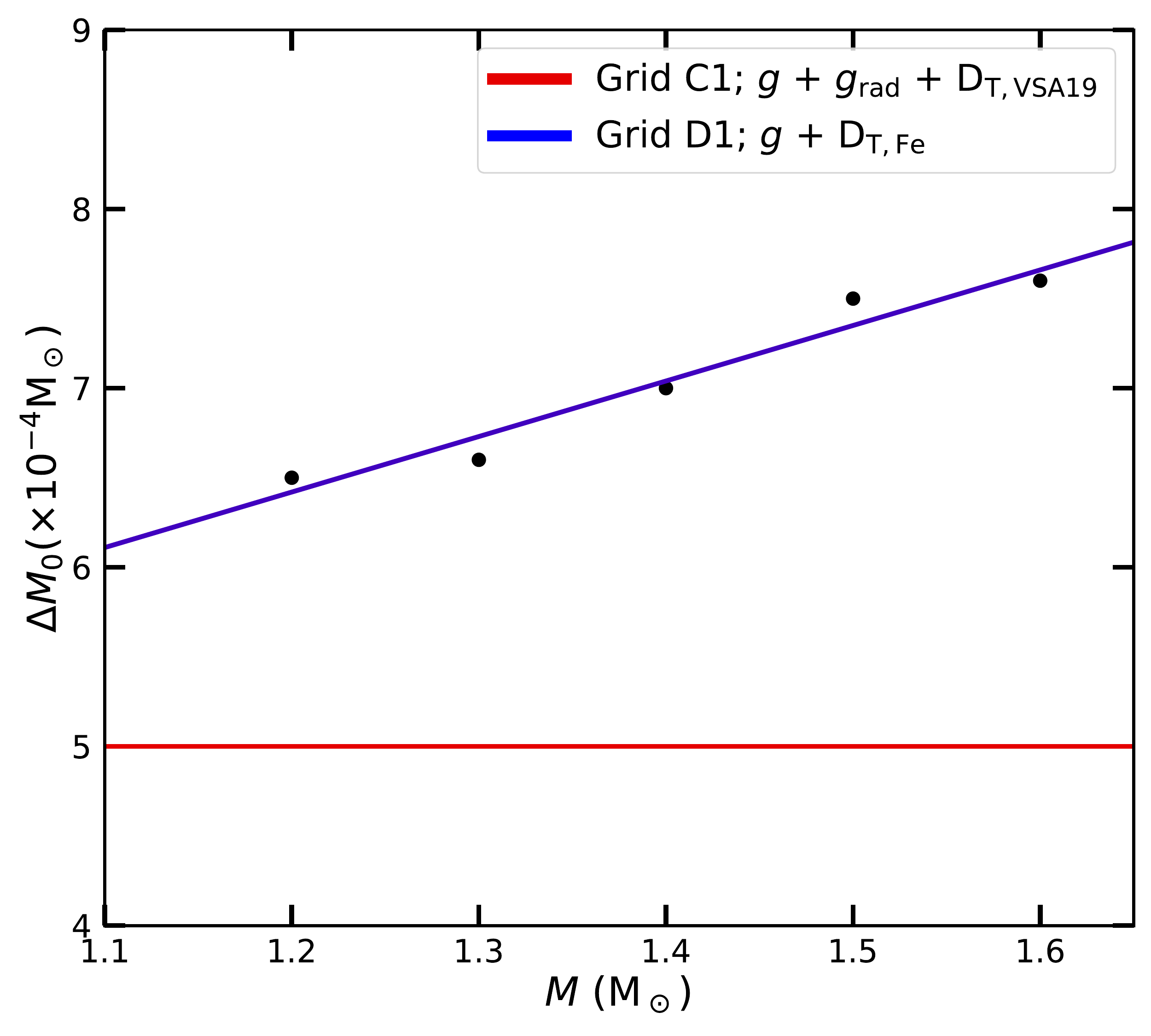}
        \caption{Values of the $\Delta M_0$ as a function of the stellar mass parametrised in Sect.~\ref{calib-dturb} (black dots) and calibrated by VSA19 (red line). The blue line represents the linear fit presented in Eq.~\ref{eq:param}. }
    \label{fig:Dturb_call}
\end{figure}

\subsubsection{Changing initial metallicity}

\begin{figure}
    \centering
    \includegraphics[width=\columnwidth]{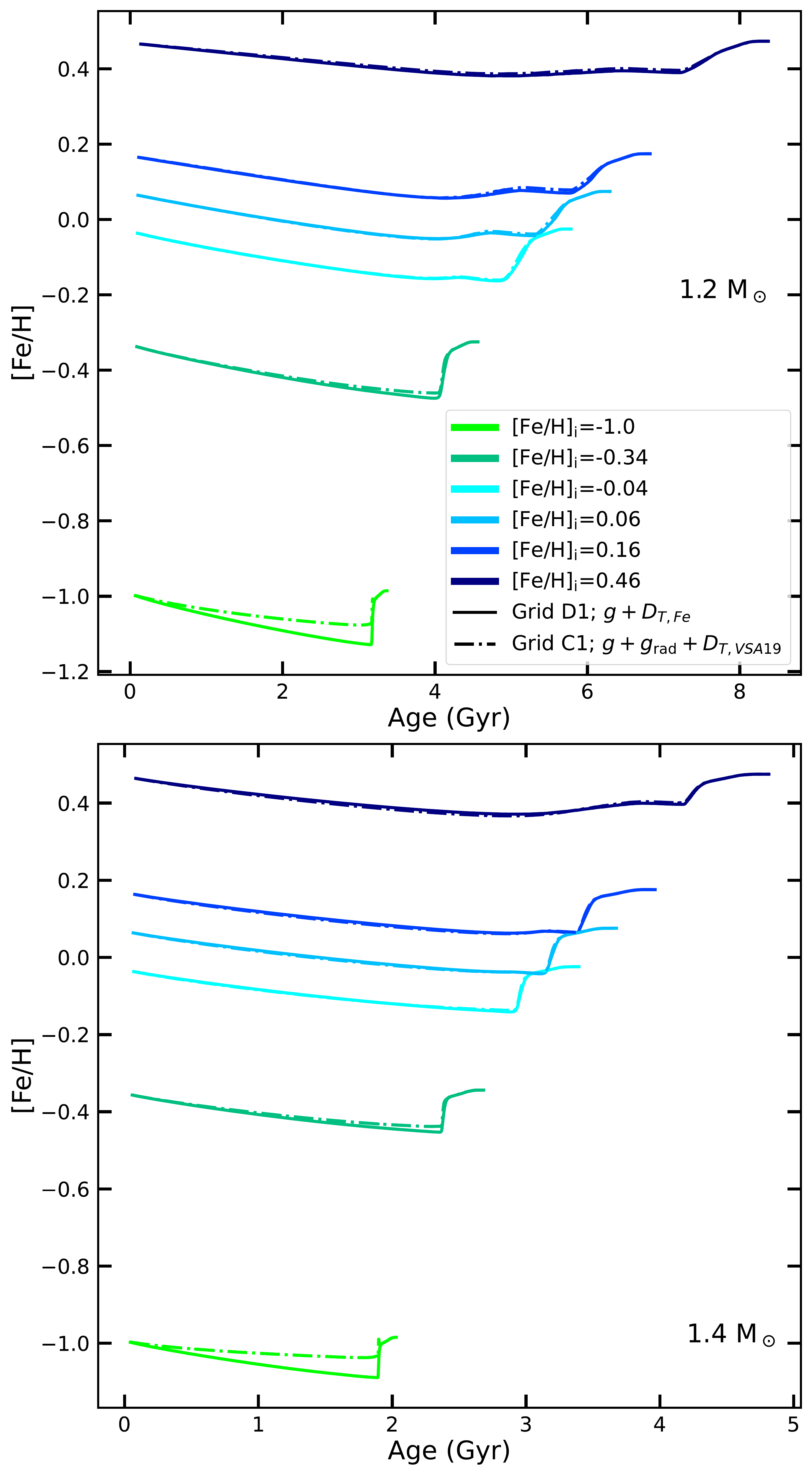}
        \caption{Evolution of [Fe/H] with time for 1.2~M$_\odot$ (top panel) and 1.4~M$_\odot$ (bottom panel) at different initial chemical compositions. The solid lines represent models including atomic diffusion (without radiative accelerations) and the parametrisation of ${D_\mathrm{T,Fe}}$ presented in Sect.~\ref{calib-dturb}, and the {dot}-dashed lines represent models including atomic diffusion with radiative accelerations and the ${D_{\mathrm{T, VSA19}}}$ calibrated by VSA19.}
    \label{fig:dMHi}
\end{figure}

As previously stated, the efficiency of atomic diffusion depends on the initial chemical composition, and so the parametrisation presented in Eq.~\ref{eq:param} may not be valid for every composition.
In order to test its validity domain, we compute, using the turbulent diffusion coefficient parametrised in the previous Section ($D_\mathrm{T,Fe}$), models with 1.2 and 1.4 M$_\odot$ and with [Fe/H]$\rm _i=$ -1.0, -0.34, -0.04, 0.06, 0.16, and 0.46~dex. Figure \ref{fig:dMHi} shows the evolution of the surface $\rm [Fe/H]$ for the different initial values.
Firstly, we see that the $D_\mathrm{T,Fe}$ models give very satisfactory iron abundance predictions for [Fe/H]$\rm _i$  between -0.4 and 0.4~dex. For lower initial metallicity, the $D_\mathrm{T,Fe}$ models deviate from models including radiative accelerations by up to 0.04~dex.
This is most likely due to the decrease in the convective envelope with [Fe/H]$\rm _i$, which allows radiative accelerations to affect the surface abundances more strongly.

\begin{figure*}[h]
    \centering
    \includegraphics[scale=0.5]{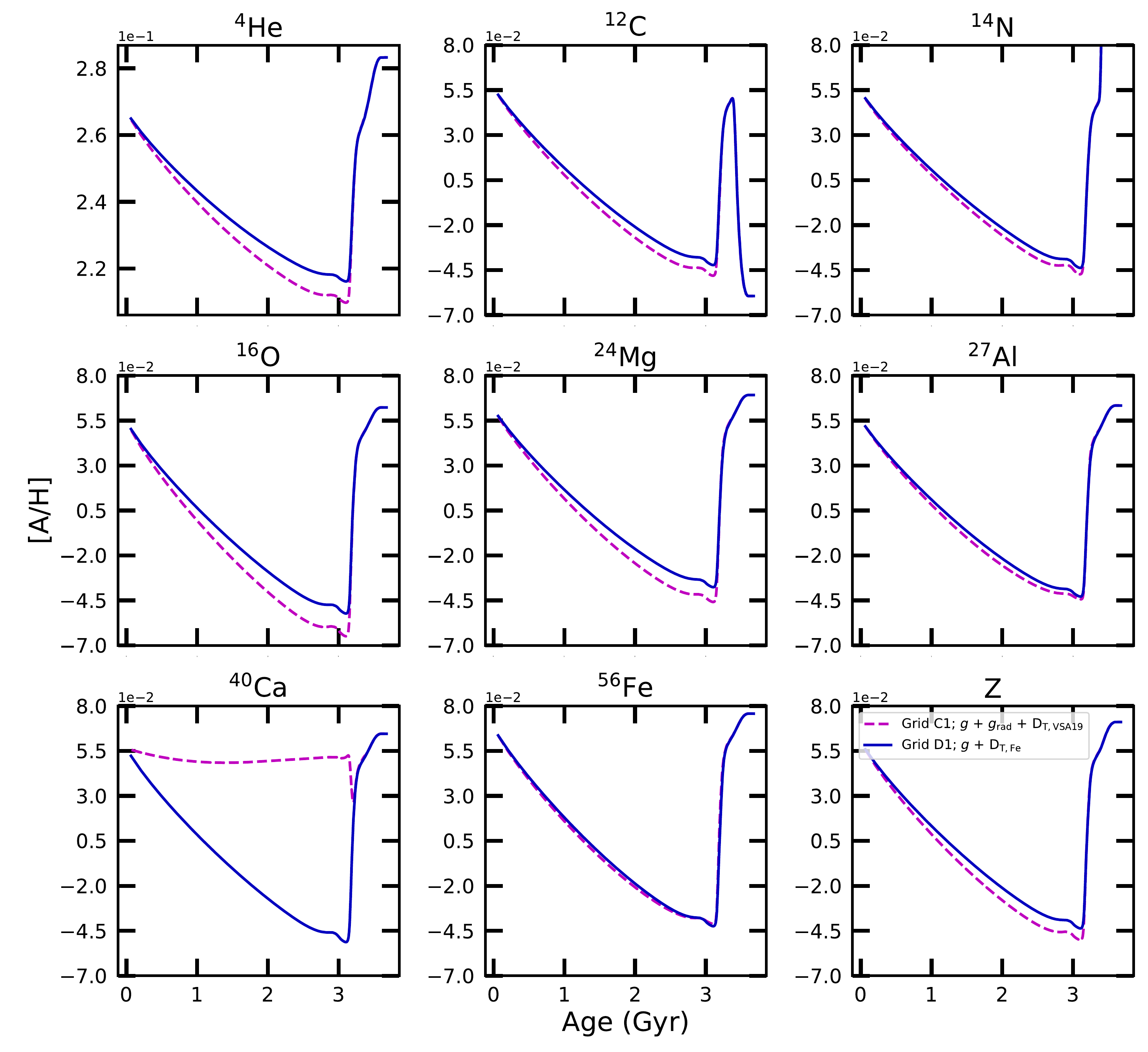}
    \caption{Evolution of the surface abundance of some  chemical elements for a 1.4~M$_\odot$ model. The {purple} dashed lines represent a model including atomic diffusion with radiative accelerations and the turbulent mixing calibrated by VSA19, and the blue solid lines represent a model including atomic diffusion without radiative acceleration and the $D_\mathrm{T,Fe}$ turbulent mixing parametrised in Sect.~\ref{calib-dturb}.}
    \label{fig:other_ele}
\end{figure*}

\subsection{Effects for other elements}
\label{diff_elem}

Each element is affected differently by radiative accelerations, and the parametrisation presented in Sect.~\ref{calib-dturb} may not be adapted for other elements than iron.
Figure \ref{fig:other_ele} shows the evolution of the surface abundances of He, C, N, O, Mg, Al, Ca, and Fe for two 1.4 M$_\odot$ models: one including turbulent mixing (${D_{\rm T,VSA19}}$) and atomic diffusion with radiative accelerations (grid C1) and the second including the parametrisation presented in this study (${D_{\rm T,Fe}}$). 

For helium (top left panel of Fig. \ref{fig:other_ele}),  the parametrised $D_\mathrm{T, Fe}$ retains more helium at the surface. However, the difference is smaller than 0.025~dex (0.006 in mass fraction), which is of the order of magnitude of the uncertainty of the helium surface abundances obtained by \cite{Verma2019a}.

Elements like carbon, nitrogen, magnesium, and aluminium (top middle, top right, middle, and middle right panel of Fig. \ref{fig:other_ele}, respectively) show good agreement with  very small differences (about 0.006~dex, 0.004~dex, 0.008~dex, and  0.004~dex, respectively). This agreement is due to the fact that these elements are either not supported by radiative accelerations (C, N, Mg) or are supported with a similar efficiency to iron (Al) in this specific case.
 
For oxygen (middle left panel of Fig. \ref{fig:other_ele}), the $D_\mathrm{T, Fe}$ parametrisation starts to show a significant difference of about 0.013~dex, which is comparable to observed uncertainties. It indicates that radiative acceleration has a smaller impact on this element compared to the others.
 
Finally, for calcium (bottom left panel of Fig. \ref{fig:other_ele}), we see that our parametrisation cannot reproduce the radiative acceleration in this element. In the model with radiative acceleration and ${D_{\rm T,VSA19}}$ turbulent mixing (Grid C1), calcium is almost at equilibrium  during the evolution (gravitational settling and radiative acceleration cancelling each other out). Radiative acceleration on calcium is greater than on iron in this case. This behaviour cannot be reproduced by  turbulent mixing calibrated on an element that is not in this situation (iron in this case), and  is the reason why the parametrisation we propose cannot be valid for all elements.

\subsection{Maximum variation in the iron surface abundance in solar-like stars}

\begin{figure}
\centering
    \includegraphics[width=0.95\columnwidth]{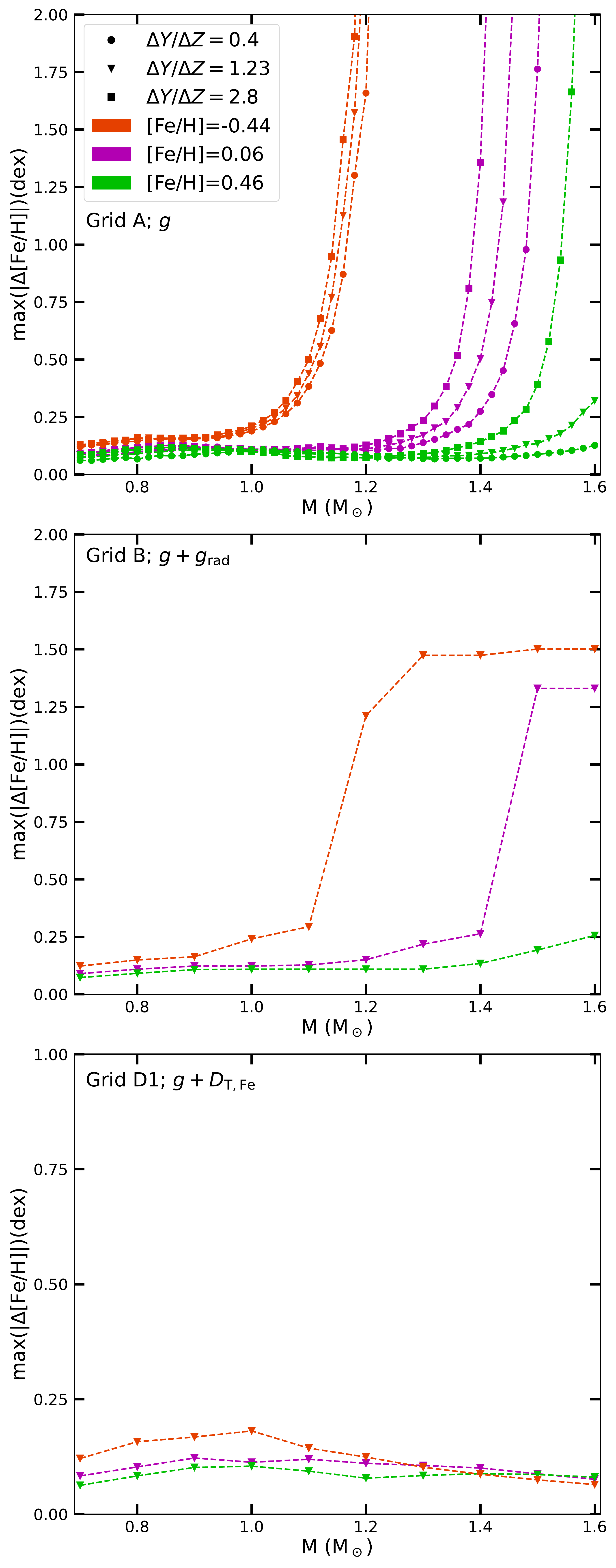}
\caption{Maximum variation in $[\rm Fe/H]$ during the evolution (up to the tip of RGB) according to  mass. The colours and symbols correspond to the values of $[\rm Fe/H]_i$ and $\Delta Y/\Delta Z$. The maximum variation was determined for ages less that 13.5 Gyr for the models that evolve on longer timescales ($M \lesssim0.95$). The top panel includes atomic diffusion without radiative accelerations, the middle panel shows models that include radiative acceleration, and the bottom panel shows $D_\mathrm{T,Fe}$ models.}
\label{fig:maxFeH}
\end{figure}

Figure \ref{fig:maxFeH} shows the maximum variation in [Fe/H] for stellar models including atomic diffusion without (top panel, Grid A) and with (middle panel, Grid B) radiative accelerations, and $D_\mathrm{T,Fe}$ models (bottom panel, Grid D1).
For models including atomic diffusion without radiative accelerations (grid A) we verify that the maximum variation (in this case the largest depletion) increases with mass as the efficiency of gravitational settling increases.
We also see that at higher masses the maximum variation reaches very high and unrealistic values (up to 2~dex) compared with variations observed in clusters \citep[e.g.][]{gruyters14,Gruyters2016,Semenova2020}.
The maximum variation also depends on the chemical composition. We see that by decreasing the initial metallicity, the variations in [Fe/H] reach unrealistic values at lower masses. Moreover, the higher  $\Delta Y/\Delta Z$ is, the sooner the depletion appears (in terms of mass).  A higher initial metallicity leads to a higher opacity, hence a higher convective region and a less efficient effect of atomic diffusion. A higher $\Delta Y/\Delta Z$ decreases the opacity, which leads to a shallower convective zone and a more efficient effect of atomic diffusion.

For models including atomic diffusion without radiative accelerations (grid B), the maximum variation in [Fe/H] corresponds to a depletion or an accumulation at the surface depending on the stellar mass. As seen in the previous plot (top panel of Fig. \ref{fig:maxFeH}), the maximum variation increases as the stellar mass increases, although for the higher stellar masses it reaches values of maximum variation that are not expected for chemically non-peculiar stars, but only for chemically peculiar stars such as Fm stars \cite[e.g.][]{Richer2000}. The fact that the maximum variation reaches a plateau is due to the saturating effect of radiative accelerations \citep{michaud15}. Changing the initial metallicity has a similar effect for models without radiative acceleration. By decreasing the initial metallicity, the variation reaches large values at smaller masses.

\begin{table*}
    \centering
        \caption[]{Global fundamental properties of the computed models and those  obtained by the optimisation.}
        \begin{tabular}{ccccccccc}
        \hline\hline
    &\multicolumn{2}{c}{Mass (M$_\odot$)}&
    \multicolumn{2}{c}{Radius (R$_\odot$)}&\multicolumn{2}{c}{Age (Gyr)}\\
    \cmidrule(lr){2-3}\cmidrule(lr){4-5}\cmidrule(lr){6-7}
    &Model& Inference &Model& Inference&Model& Inference\\\hline
    
    \multicolumn{1}{c}{Model 1}&1.2&$1.20\pm0.01$&1.240&$1.240\pm0.004$&2.1&$2.1\pm0.2$ \\
    
   \multicolumn{1}{c}{ Model 2}&1.4&$1.40\pm0.01$&1.598&$1.596\pm0.004$&1.7&$1.7\pm0.1$\\
    
    \multicolumn{1}{c}{Model 3}&1.6&$1.60\pm0.01$&1.870&$1.868\pm0.004$&1.2&$1.2\pm0.1$\\\hline
    
\end{tabular} 
        \label{table:mod_op}
\end{table*}

For the $D_\mathrm{T,Fe}$ models (grid D1), we see that the maximum variation in iron is smaller than 0.2 dex during the whole evolution from the  MS to RGB. As expected, this shows that turbulent mixing avoids the unrealistic chemical abundance variations induced by atomic diffusion.

\section{Impact on the stellar properties inference}
\label{sec:stars_infer}

In this section we test the models including the parametrisation of turbulent mixing $D_\mathrm{T,Fe}$ using classical and seismic constraints. For the optimisation of the stellar fundamental properties, we use the code Asteroseismic Inference on a Massive Scale  (AIMS; \citealt{Rendle2019}).

\begin{figure*}[h]
    \centering
    \includegraphics[scale=0.35]{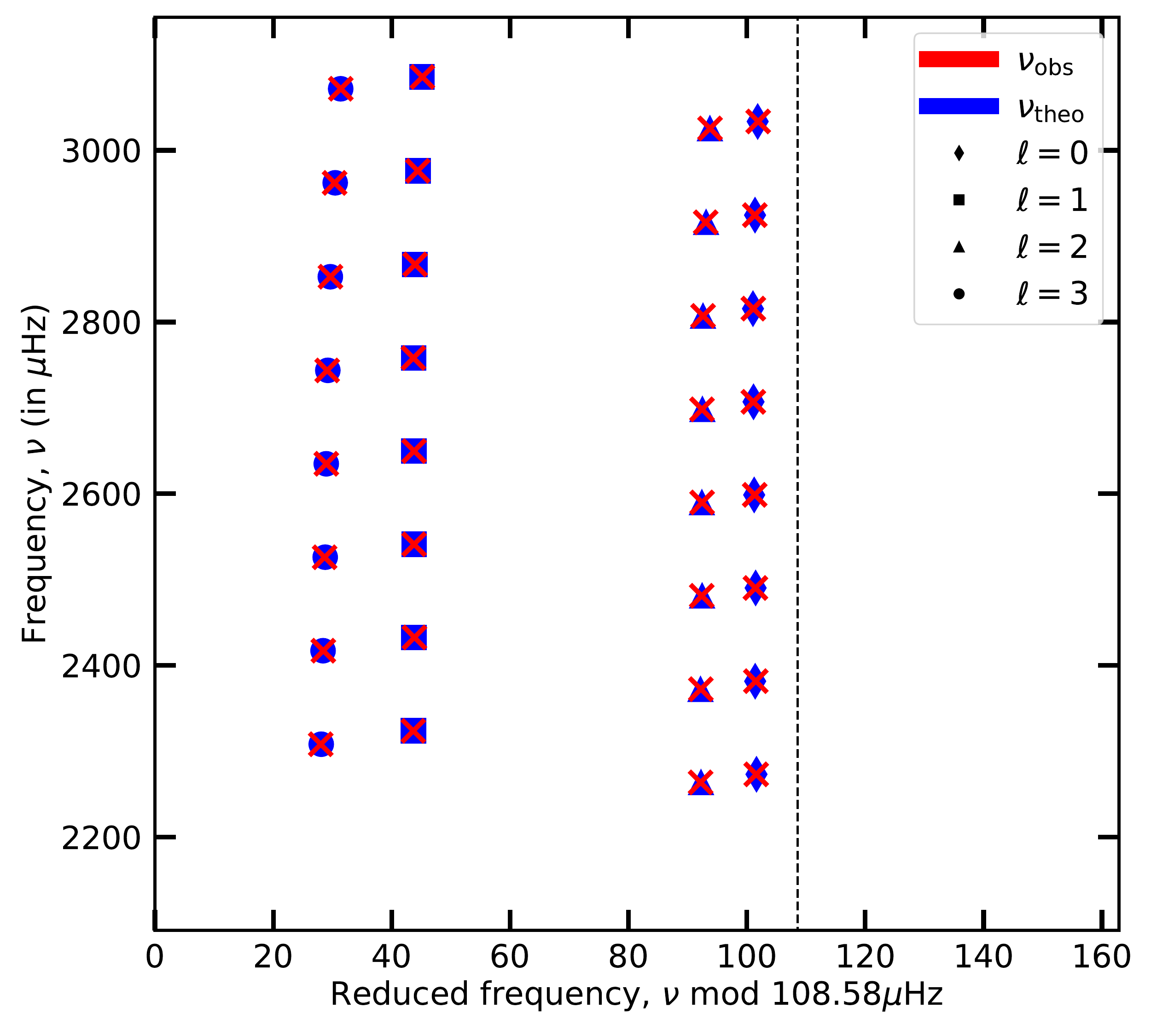}
    \includegraphics[scale=0.35]{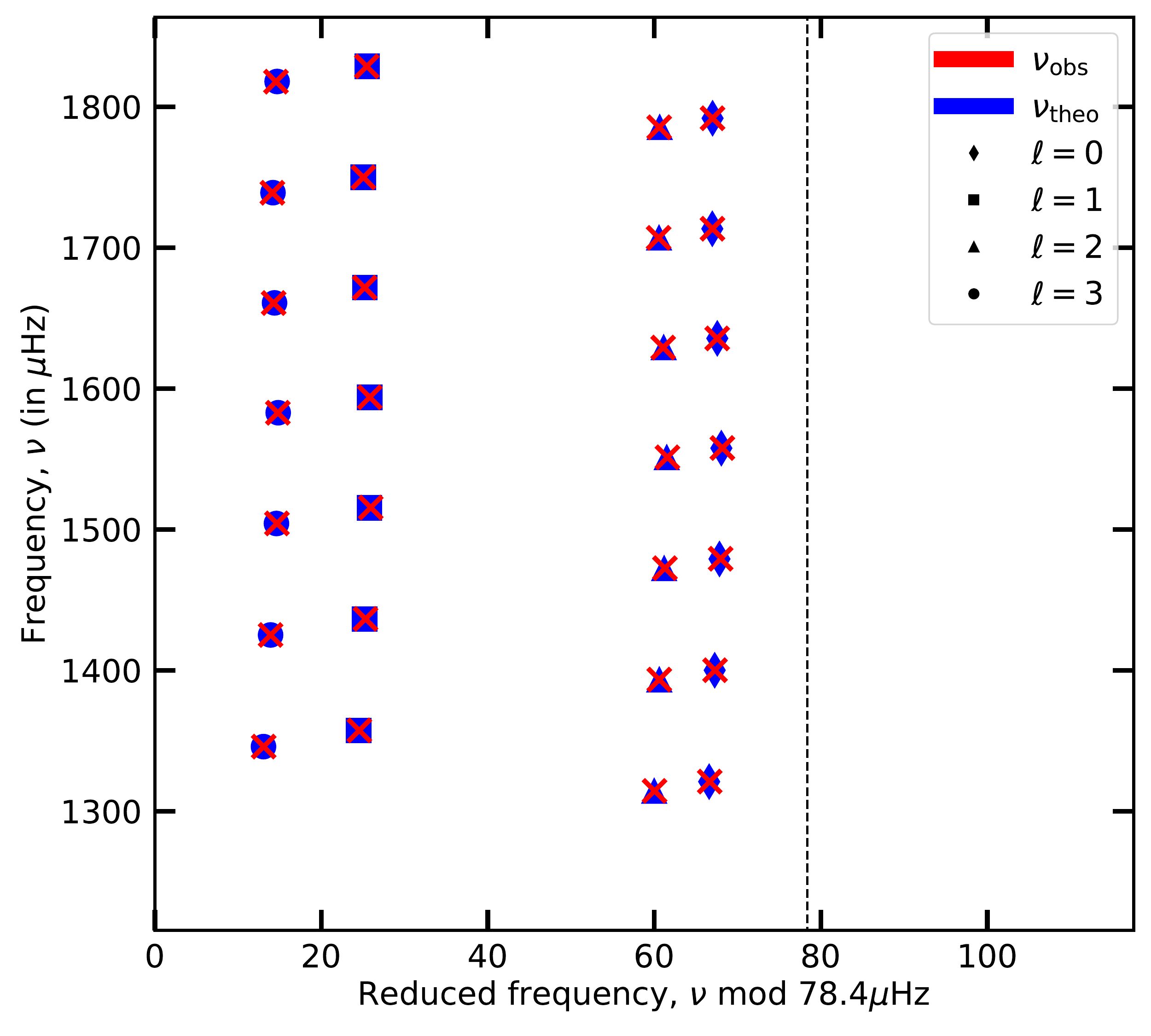}
    \includegraphics[scale=0.35]{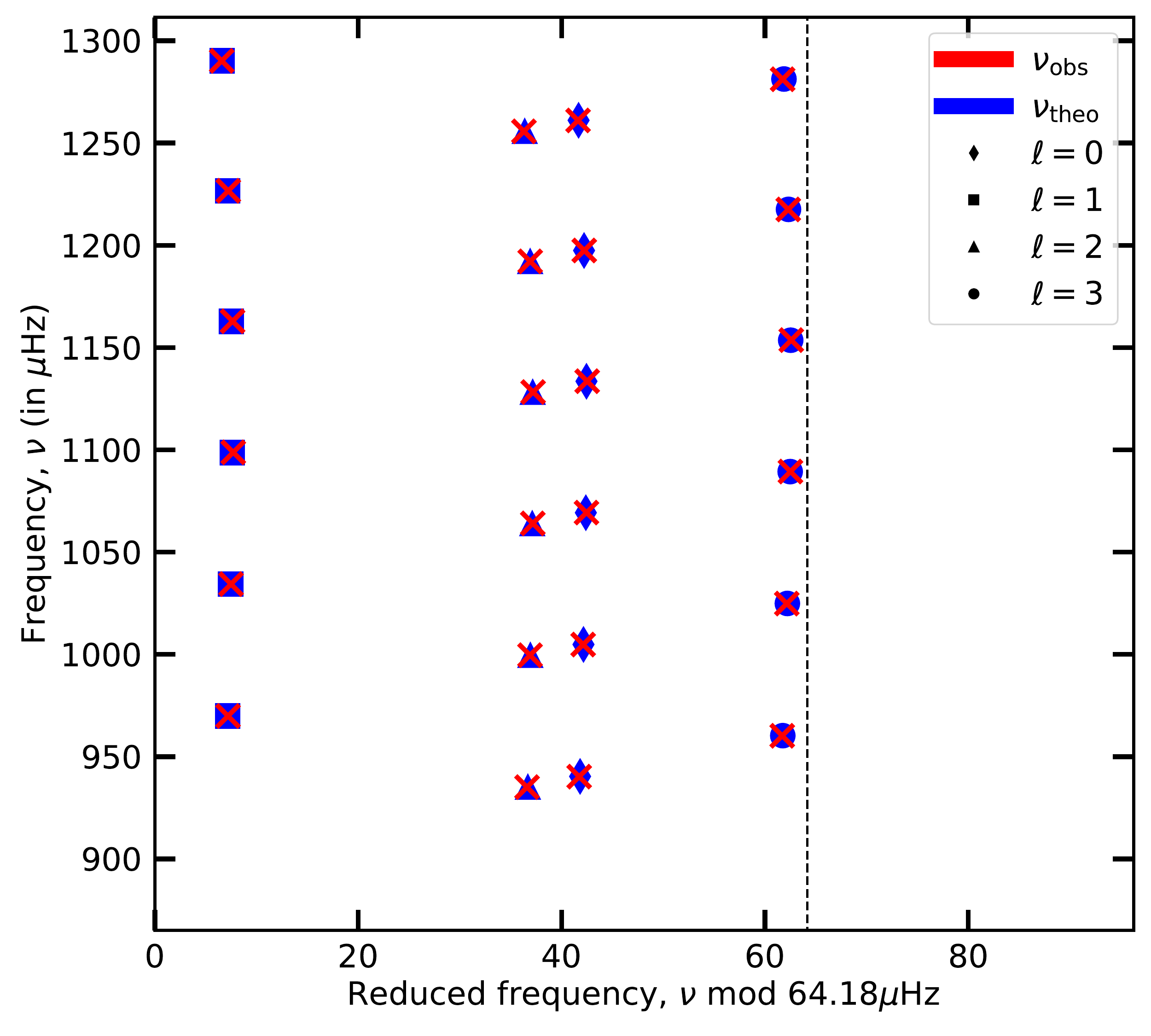}
    \caption{Echelle diagram of the best optimisation result.  The top panel is for a star with 1.2~M$_\odot$, the middle is for 1.4~M$_\odot$, and the bottom is for 1.6M$_\odot$. ${\nu_\mathrm{obs}}$ are the frequencies used as constraints from the optimised model from Grid D1; ${\nu_\mathrm{theo}}$ are the optimised frequencies from Grid C1.}
    \label{fig:freq_t}
\end{figure*}

\subsection{Seismic validation of the models}\label{AIMS_models}

To test the seismic validity of the parametrisation, we select three models along the tracks with masses equal to 1.2, 1.4, and 1.6 M$_\odot$ and  [Fe/H ]$\rm_i=0.06~dex$ computed including atomic diffusion without radiative accelerations and the turbulent mixing calibrated in Section~\ref{calib-dturb} ($D_\mathrm{T,Fe}$), as targets (from Grid D1).
Then we derive their global fundamental properties (mass, radius and age) using the AIMS optimisation code and the grid {C1} (see Section~\ref{sec:stellar_models}), in order to quantify the validity of the parametrisation. We use the effective temperature, [Fe/H], and the individual frequencies as constraints. We give the two classical constraints the same weight as all the individual frequencies. Because the targets are models, the surface correction terms are not applied. The results are shown in Table~\ref{table:mod_op} where we see that the properties of the three target models are well retrieved within 1$\sigma$. Figure \ref{fig:freq_t} shows that the echelle diagrams of the best fitting models of the grid are overlapping one of the three targets. This indicates that the parametrised models give similar results in terms of seismic frequencies to the models including radiative accelerations and can be { used to infer the stellar properties.}

\subsection{Application on \textit{Kepler} stars}

We applied the $D_\mathrm{T,Fe}$ models for the inference of the global properties of two stars from the Kepler Legacy sample, KIC~2837475 and KIC~112253226. These are two of the three stars used in VSA19. We selected these two stars because they have similar chemical compositions, which  allowed us to build a smaller grid for the optimisation procedure. All the seismic and classical constraints are taken from \cite{Lund2017}. We also considered the two  surface correction terms of \citet{Ball2014}. Table \ref{table:star_op} presents the results of the inferred global properties using different grids of stellar models. 

We used grids D1, D2, D3, and D4 in order to disentangle the impact of atomic diffusion and the turbulent mixing on the inferred fundamental properties. The comparison of the results obtained with grids D1 and D2 shows the impact of both processes compared to a grid without transport (except convection) using the same solar calibration as grid D1. In this case the masses and radii are similar with both grids and for the two stars. The main impact of including transport in the models is on the age (about 11\%\ and 12\% for KIC~2837475 and KIC~11253226, respectively).

The comparison of the results obtained with grids D1 and D3 shows the same impact but with both grids being calibrated to the Sun according to their input physics. The age differences are slightly larger, about 13\%\ and 15\% for KIC~2837475 and KIC~11253226, respectively. For the radius the difference is 2\% for KIC~2837475, and for the mass the differences are about 1\% for both stars. Nevertheless, masses and radii are all within 1$\sigma$.

The comparison of grids D1 and D4 shows an estimation of the impact of atomic diffusion alone close to the core. To do this we fitted the observed [Fe/H] with the initial [Fe/H] of the models in order not to include the effect of the unrealistic depletion of iron at the surface when no competing transport processes are taken into account. Again, both grids are calibrated to the Sun according to their input physics. The differences in mass and radius are around 1\% at maximum. The main effect is on age with similar values to the comparisons with grids D2 and D3. This seems to indicate that most of the effect on age is due to atomic diffusion (mainly gravitational settling close to the core), while turbulent mixing mainly affects the mass and radius determination by inducing a realistic [Fe/H] at the surface of the models. However, we cannot take this as a strong conclusion considering the large error in age for the grid D4.

\begin{table}
        \caption[]{Global fundamental properties obtained for the two studied stars.}
        \resizebox{\columnwidth}{!}{%
        \begin{tabular}{ccccccccc}
        \hline \hline

KIC      & Grid            & Mass (M$_\odot$) & Radius (R$_\odot$) & Age (Gyr)     \\ \hline
2837475  & D1              & 1.43 $\pm$ 0.03  & 1.64$\pm$0.01      & 1.54$\pm$0.11 \\ 
2837475  & D2              & 1.42 $\pm$ 0.03  & 1.64$\pm$0.01      & 1.71$\pm$0.12 \\ 
2837475  & D3              & 1.40$\pm$0.03    & 1.62$\pm$0.01      & 1.74$\pm$0.12 \\ 
2837475  & D4              & 1.40$\pm$0.05    & 1.64$\pm$0.02      & 1.82$\pm$0.32\\ \hline
11253226 & D1              & 1.41 $\pm$0.02   & 1.61$\pm$0.01      & 1.53$\pm$0.11 \\ 
11253226 & D2              & 1.41 $\pm$ 0.03  & 1.61$\pm$0.01      & 1.71$\pm$0.11 \\ 
11253226 & D3              & 1.39$\pm$0.03    & 1.61$\pm$0.01      & 1.76$\pm$0.11 \\ 
11253226 & D4              & 1.41$\pm$0.04    & 1.61$\pm$0.02      & 1.60 $\pm$0.21 \\ \hline
\end{tabular} }
        \label{table:star_op}
\end{table}

\section{Conclusions}
\label{conclu}

In this work, we used the MESA stellar evolutionary code to compute all the stellar models. We first quantified the effect of atomic diffusion, with and without radiative accelerations, on the evolution of the surface abundance of iron. We confirmed that our models are consistent with previous studies. We also showed that [M/H] and [Fe/H] can be different ($\sim 0.02$~dex depending on the mass and the transport processes of chemical elements included in the models, i.e. comparable with the error on iron abundance). The observed [Fe/H] being the main constraint on the chemical abundance of stars, comparing it with model predictions of [M/H] may then lead to large uncertainties, especially for the determination of stellar fundamental properties.

We also quantified the effects of turbulent mixing in stellar models. We first considered the calibration done by VSA19. As expected, turbulent mixing is efficient at preventing strong variations in the surface abundances induced by atomic diffusion. Nevertheless, it does not suppress the effect of atomic diffusion. In the context of accurate stellar fundamental property inferences (of mass, radius, and age)  of large samples of stars, there is a need for large grids of stellar models including these processes. Because the computation of large grids including atomic diffusion with radiative acceleration is computationally expensive, we propose a parametrisation of the turbulent mixing to include the competing effect of radiative acceleration. We focused the parametrisation on iron alone because this is currently the only element used as a chemical constraint. To achieve this, we parametrised an increase in the efficiency of turbulent mixing to match the competition of the radiative acceleration. For the other elements the parametrisation performs rather well, except for oxygen and calcium. For helium the difference induced by the parametrisation is of the same order of magnitude as the uncertainties on the helium surface abundances obtained from \textit{Kepler} stars \citep{Verma2019a}. However, this parametrisation should be used for studies relying only on iron abundances. For a better chemical characterisation, a full treatment of radiative acceleration should still be preferred.

We finally compared models including the full treatment of atomic diffusion and turbulent mixing with the $D_\mathrm{T,Fe}$ models. We found no relevant differences in the seismic properties. We also performed the characterisation of two F-type stars of the \textit{Kepler} Legacy sample (KIC~2837475 and KIC~11253226) with four different grids (D1, D2, D3, and D4). We identified that the main effect is on age reaching differences of about 11--15\% between the $D_\mathrm{T,Fe}$ models and models without transport except for convection.  We also identified that atomic diffusion is the main contributor to this difference in age. However, we cannot draw a strong conclusion considering the possible cancellation effect of atomic diffusion and turbulent mixing.

The proposed parametrisation makes the computation of large grids of stellar models as fast as those including atomic diffusion without radiative acceleration. Nonetheless, the parametrisation should be performed any time the input physics is different. For example, input physics predicting hotter models would require a more efficient turbulent mixing to account for the more efficient radiative accelerations.

The parametrisation was designed especially for the purposes of fundamental property inference, and for stellar evolution codes including time-consuming atomic diffusion (with radiative acceleration) computation. The implementation of such parametrisation is straightforward and can be applied using the `hook' functionality of MESA. The natural next step is the implementation of the single valued parameter approximation in MESA in order to compute models with radiative accelerations in a more efficient way, which requires a deeper modification of the code.
This work is only a first step towards  the computation of large grids of stellar models including the best compromise between accuracy and computational cost.

\section*{Acknowledgements}

This work was supported by FCT/MCTES through the research grants UIDB/04434/2020, UIDP/04434/2020 and PTDC/FIS-AST/30389/2017, and by FEDER - Fundo Europeu de Desenvolvimento Regional through COMPETE2020 - Programa Operacional Competitividade e Internacionalização (grant: POCI-01-0145-FEDER-030389).  NM acknowledges support from the Fundação para a Ciência e a Tecnologia (FCT) through the Fellowship UI/BD/152075/2021 and POCH/FSE (EC). DB and MD are supported by national funds through FCT in the form of a work contract. We thank the anonymous referee   for the valuable comments which helped to improve the paper. We also thank Elisa Delgado-Mena for fruitful discussion.

\bibliographystyle{aa} 
\bibliography{references.bib} 
\newpage\phantom{---}
\newpage

\begin{appendix}

\section{Evolution of the iron surface abundance for different transport processes of chemicals}\label{appendix:A}

\begin{figure*}
\centering
    \includegraphics[width=1.8\columnwidth]{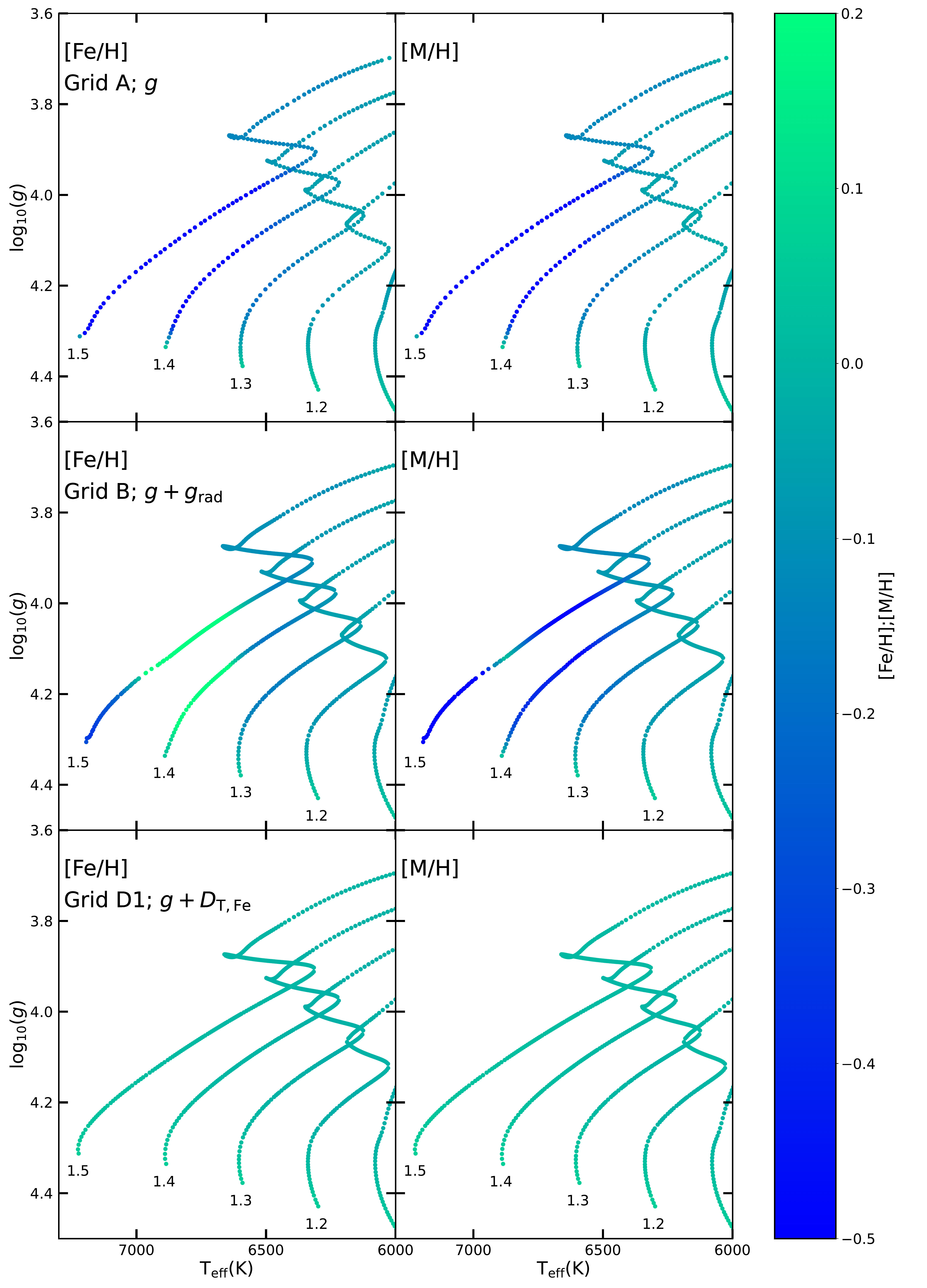}
    \caption{Kiel diagram of stars with [Fe/H]$\rm _i$ and $\frac{\Delta Y}{\Delta Z}=1.23$ (see description in   Appendix~\ref{appendix:A}). The sub-panels are labelled as follows: {$g$ for models that include atomic diffusion without radiative accelerations; $g+g_\mathrm{rad}$ for  models that include atomic diffusion with radiative accelerations; $g+D_\mathrm{T,Fe}$ for models that include atomic diffusion without radiative accelerations and the turbulent diffusion coefficient parametrised in Section~\ref{calib-dturb} (see Table~\ref{table:stellar_grids} for more details about the input physics).}}
    \label{fig:mdiagr}
\end{figure*}

In this appendix we present Fig. \ref{fig:mdiagr}, which shows a Kiel diagram similar to Fig. \ref{fig:HR} for some evolutionary tracks focused on the MS of stars with masses higher than 1.2 M$_\odot$. The top panels show models from Grid A, the middle panels from  Grid B, and the bottom panels  from Grid D1. For the left panels, the colour represents the surface [Fe/H], while it represents the [M/H] for the right panels. 

For the [Fe/H], in the case with only gravitational settling (top panel) we see the depletion of the surface abundances due to atomic diffusion, which is  stronger for the higher mass stars. When the radiative acceleration is included (middle panel) we see that an accumulation effect can occur
 for stars with 1.4 M$_\odot$ or higher, instead of the depletion effect for the smaller ones. For models with turbulent mixing (bottom panels), the variations caused by atomic diffusion are smaller and almost unnoticeable in this diagram.
For the case of [M/H] (right panels), the results are similar to those for [Fe/H], except for the case with radiative accelerations where we see depletion instead of enrichment in the surface abundances.

\nopagebreak
\FloatBarrier

\section{Calibrating the turbulent mixing}
\label{sec:calibr_dturb}

\begin{figure*}
\centering
    \includegraphics[width=1.8\columnwidth]{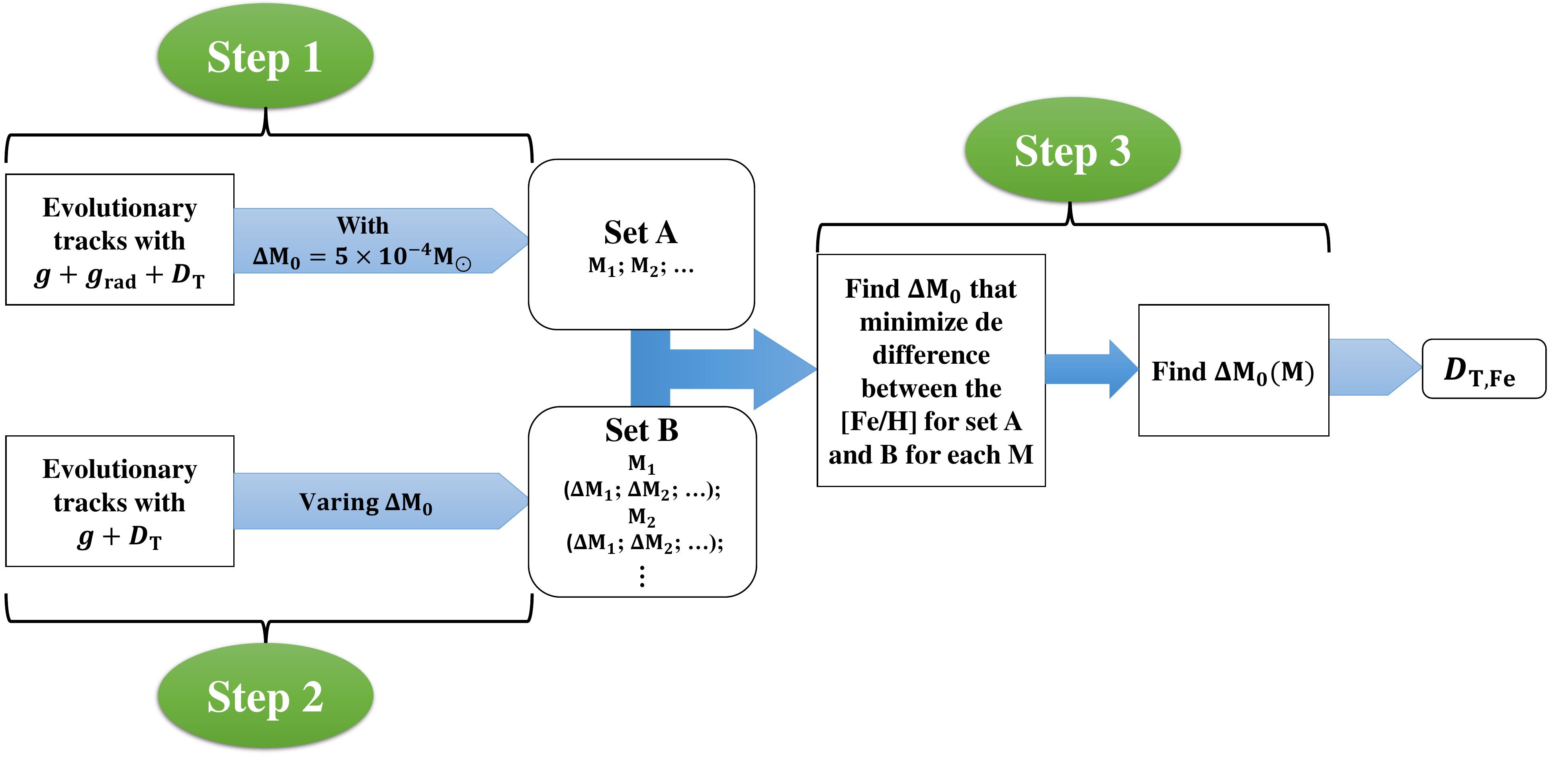}
    \caption{Parametrisation procedure of $\Delta M_0$.}
    \label{fig:scheme}
\end{figure*}

The parametrised turbulent mixing proposed in this study depends on the stellar physics used in the models. Therefore, it is necessary to calibrate the reference mass any time the physics is changed. We provide here the general steps used to recalibrate our parametrisation for any physics considered:

\begin{itemize}
    \item[Step 1:] Compute the reference stellar models, with different masses $\geq$ 1.2 M$_\odot$, including atomic diffusion with radiative accelerations, turbulent mixing with the $\Delta M_0$ value of choice, and the solar initial chemical composition obtained from a solar calibration (set A).
    
    \item[Step 2:] For each reference model from Step 1, compute a new set of models (set B) including atomic diffusion without radiative acceleration, and different values of $\Delta M_0$.
    
    \item[Step 3:] Compare the surface [Fe/H] evolution of the reference models with those computed with different $\Delta M_0$ values in Step 2 for every mass of the grids. The best value is obtained by a minimisation procedure. The relation between $\Delta M_0$ and the mass is then obtained with linear regression.
    
\end{itemize}

\end{appendix}

\end{document}